\documentclass[11pt,a4paper]{article}
\usepackage{geometry}                

\usepackage{nicefrac}

\usepackage{jheppub}

\usepackage{amsmath,amssymb,graphicx} 

\usepackage{psfrag}
\usepackage{subfigure}
\usepackage{tabularx}
\usepackage{footmisc}

\newcommand\cc[1]{#1^{^{\kern-6pt \circ}}\kern2pt}

\def\pa{\partial}
\renewcommand{\a}{\alpha}
\renewcommand{\b}{\beta}

\newcommand{\m}{\mu}
\newcommand{\n}{\nu}

\def\be{\begin{equation}}
\def\ee{\end{equation}}
\def\bea{\begin{eqnarray}}
\def\eea{\end{eqnarray}}
\def\ba{\begin{array}}
\def\ea{\end{array}}
\def\bi{\begin{itemize}}
\def\ei{\end{itemize}}

\newcommand{\beq}{\begin{equation}}
\newcommand{\eeq}{\end{equation}}
\newcommand{\beqn}{\begin{eqnarray}}
\newcommand{\eeqn}{\end{eqnarray}}
\newcommand{\bga}{\begin{align}}

\def\dalemb#1#2{{\vbox{\hrule height .#2pt
\hbox{\vrule width.#2pt height#1pt \kern#1pt
\vrule width.#2pt}
\hrule height.#2pt}}}

\title{\Huge Holographic fluids with vorticity and analogue gravity}
\preprint{CPHT-RR002.0112}
\author[a,b]{\bf Robert G. Leigh,}
\author[c]{\bf Anastasios C. Petkou,}
\author[d]{\bf and P. Marios Petropoulos}
\affiliation[a]{Department of Physics, University of Illinois\\1110 W. Green Street, Urbana IL 61801, U.S.A.}
\affiliation[b]{Perimeter Institute for Theoretical Physics, \\31 Caroline Street North, Waterloo ON, Canada  N2L 2Y5}
\affiliation[c]{Department of Physics, University of Crete\\71003 Heraklion, Greece}
\affiliation[d]{Centre de Physique Th\'eorique, Ecole Polytechnique, CNRS UMR 7644\\91128 Palaiseau Cedex, France}
\emailAdd{rgleigh@illinois.edu}
\emailAdd{petkou@physics.uoc.gr}
\emailAdd{Marios.Petropoulos@cpht.polytechnique.fr}
\date{\today}                                          

\begin{document}

\abstract{
We study holographic three-dimensional fluids with vorticity in local equilibrium and discuss their relevance to analogue gravity systems.  The Fefferman--Graham expansion leads to the fluid's description in terms of a comoving and rotating Papapetrou--Randers frame. A suitable Lorentz transformation brings the fluid to the non-inertial Zermelo frame, which clarifies its interpretation as moving media for light/sound propagation.  We apply our general results to the Lorentzian Kerr--AdS$_4$   and Taub--NUT--AdS$_4$  geometries that describe fluids in cyclonic and vortex flows respectively. In the latter case we associate the  appearance of closed timelike curves to analogue optical horizons. In addition, we derive the classical rotational Hall viscosity of three-dimensional fluids with vorticity.  
 Our formula remarkably resembles the corresponding result in magnetized plasmas. }
\maketitle
\section{Introduction and motivation}

The fluid/gravity correspondence advocates that the long-distance hydrodynamical degrees of freedom of a  quantum system residing at the boundary of an asymptotically AdS space are holographic images of  gravitational degrees of freedom in the bulk. This is a novel paradigm for emergence in physics which, not surprisingly, has attracted the interest of the  physics community outside high energy, since it potentially provides new insight into the long-distance physics of diverse strongly coupled systems.  

One of the first applications of the above reasoning was in the physics of the quark-gluon plasma\footnote{See for example \cite{Janik} for a recent review.}.  Much wider in scope is the recent  interest in the AdS/CMT correspondence, namely the efforts to apply  holographic techniques to a number of strongly coupled condensed matter systems. In principle, the landscape of the latter offers an almost inexhaustible number of potential applications for AdS/CMT. In practice, few systems have been comprehensively studied so far, such as high-T$_c$ superconductors \cite{HHH}, quantum Hall fluids \cite{Esko}, non-Fermi liquids \cite{Liu} and more recently topological insulators \cite{Hoyos,HLF}. 

Simultaneously, relativistic hydrodynamics of both normal fluids and superfluids  in $3+1$ dimensions has been extensively analyzed and compared to gravitational dynamics in $4+1$ dimensions (for a review see \cite{Rangamani}). Recently the interest turned to $2+1$ dimensions and to parity non-preserving hydrodynamics which has been studied in \cite{Kovtun, Minwalla} focusing on the classification  of the different transport coefficients. Interesting explicit holographic results on anomalous transport coefficients have also lately appeared \cite{Son,Landsteiner}.   

The main motivation behind our work is to extend the realm of  AdS/CMT to two large classes of systems of considerable current interest to the condensed matter community. The first class includes fast rotating atomic gases \cite{Cooper,Fetter,Dalibard} as it is anticipated that new experimental techniques will allow one to reach the strong-coupling regime in which the ratio of the number of vortices to number of particles is of order 1.  In such a case a ``bosonic fractional quantum Hall state'' might be observed. The second class contains the so-called analogue gravity systems (for a comprehensive review see \cite{Barcelo}). The crucial underlying physics behind these systems is sound/light propagation in supersonically/superluminally (with respect to the local velocity of sound/light) moving media. This is currently achieved using various meta-materials with dedicated acoustic/optical properties. The supersonically/superluminally moving regions are bounded by acoustic/optical analogue horizons. Phenomena similar to Hawking radiation 
for phonons/photons could possibly appear in the vicinity of these horizons  \cite{Unruh:1980cg, Unruh:1994je} and are experimentally investigated \cite{Cacciatori,Liberati}.

The common base for the possible description of the above systems would be  holographic fluids/superfluids in local thermal equilibrium that have non-trivial kinematic properties, and in particular vorticity. Once we establish such systems, we can then try to calculate their transport properties studying fluctuations and using Kubo-like formulas.
Although a flowing fluid generically dissipates, we expect that the fluids at the boundary of 
time-independent bulk solutions are 
stationary
and their energy--momentum tensor has just the perfect relativistic-fluid form. Since the present work aims to initiate the holographic description of rotating atomic gases and analogue gravity systems, we will focus on flows having vorticity, but zero shear, expansion and acceleration. In that sense our boundary systems are the neutral analogues of  Hall fluids and hence could be interpreted as rotating neutral gases. Furthermore, studying sound/light propagation through such boundary fluids should be equivalent to a holographic study of analogue gravity systems\footnote{A first attempt at a holographic description of acoustic analogue gravity systems was made in \cite{Shapere}.}.

Our starting point is the holographic $3+1$-split formalism \cite{Leigh:2007wf,Mansi:2008br,Mansi:2008bs} and the corresponding Fefferman--Graham  expansion. The latter gives rise to two tensor structures -- the boundary metric and the stress tensor of the boundary fluid, which are in principle independent. Equivalently, we can think of these data as a frame and a velocity field $\hat{u}$. We show that the Fefferman--Graham expansion corresponds to choosing the comoving, and generally rotating, Papapetrou--Randers  \cite{Gibbons} frame for the fluid's observer. Then, we show that a local Lorentz transformation leads to the description of the boundary fluid using the non-inertial  Zermelo frame. The latter should not be confused with the ZAMO\footnote{Zero Angular Momentum Observer.} frame used in astrophysics when one wants to eliminate inertial forces from the observation frame \cite{Bardeen}. In some instances, however, Zermelo and ZAMO frames do coincide, as we will see on a specific example.

The Zermelo frame description clarifies the boundary physics as being that of a moving fluid.
This is the appropriate set up for the holographic description of both rotating atomic gases and analogue gravity systems. In particular,  the Zermelo frame metric has the typical form of an analogue spacetime used for sound \cite{Barcelo} and light \cite{Leonhardt} propagation in moving media. This way we can identify acoustic/optical analogue horizons in our boundary fluids. 

Our explicit examples are the Kerr--AdS$_4$ (KAdS) and Taub--NUT--AdS$_4$ (TNAdS) exact solutions and their boundary fluids. We show that they describe holographically a conformal fluid in cyclonic and vortex flow respectively. We analyze in detail the case of TNAdS whose boundary metric possesses closed timelike curves (CTCs), as a direct consequence of the homogeneous vorticity it carries. We discuss this facet of homogeneity and explain why the CTCs do not affect the classical motion of the fluid's elements. These properties are \emph{in fine} related to the absence of a globally defined spacelike constant-time surface with respect to the comoving observer. For Zermelo observers, this has the consequence that the 
fluid's elements near the vortex are dragged into superluminal rotation. The onset of this phenomenon defines an edge, which we interpret as an analogue horizon for light propagation in our $2+1$-dimensional fluid, that is, an optical horizon. The above fluids also exhibit  acoustic horizons, which however we do not further study in this work.

In our approach to holographic fluids we have clarified the important role of the different observer's frames in a particular background geometry. As an important spinoff of our understanding, we derive a simple formula for the {\it rotational Hall viscosity}, which is the non-dissipative transport coefficient that multiplies the momentum flow generated by small fluctuations of the background geometry in the presence of vorticity.\footnote{We were not able to find our result in the recent works on parity non-invariant hydrodynamics \cite{Kovtun,Minwalla}.}  This is the gravitational analogue of the classical Hall conductivity and it is remarkably similar to the {\it classical Hall viscosity} coefficient for anomalous momentum transport in magnetized plasmas \cite{Read}.

The organization of this work is as follows. In Sec. \ref{holflu} we discuss the holographic description of fluids with non-zero vorticity, their Papapetrou--Randers and Zermelo frames and their interpretation as analogue gravity systems. In Sec. \ref{geom} we review the properties of Kerr--AdS$_4$ (KAdS) and Lorentzian Taub--NUT--AdS$_4$ (TNAdS) geometries. In Sec. \ref{KTNfl} we discuss the dissipationless transport of three-dimensional relativistic fluids with vorticity and present the corresponding  Kubo formula. Section \ref{visc} contains our conclusions and perspectives.

\section{Stationary holographic fluids with vorticity} \label{holflu}

This section is devoted to the description of the formalism we will use, namely the Fefferman--Graham (FG) expansion along the holographic radial coordinate. Anticipating the specific results on Kerr--AdS and Taub--NUT--AdS presented  in Sec. \ref{KTNfl}, we discuss some generic properties of the boundary geometries and the hosted holographic fluids. Emphasis is given to the energy--momentum tensor and its dissipative components as well as on the appearance of Papapetrou--Randers and Zermelo frames. The relationship with analogue gravity systems is also set here.

\subsection{The $3+1$-split formalism and the Fefferman--Graham expansion}
We find  illuminating  to discuss holographic fluid dynamics starting from the $3+1$-split formalism introduced in \cite{Leigh:2007wf, Mansi:2008br, Mansi:2008bs}. We begin with the Einstein--Hilbert action in the Palatini first-order formulation
\bea
\label{HPaction}
S&=&-\frac{1}{32\pi G_\mathrm{N}}\int \epsilon_{ABCD}\left(R^{AB}-\frac{\Lambda}{6}E^A\wedge E^B\right)\wedge E^C\wedge E^D\nonumber \\
&=& \frac{1}{16\pi G_\mathrm{N}}\int \mathrm{d}^4x\sqrt{-g}(R-2\Lambda)\,,
\eea
where $G_\mathrm{N}$ is Newton's constant. We also assume negative cosmological constant expressed as $\Lambda =-\nicefrac{3}{L^2}=-3k^2$. 
We denote the orthonormal coframe $E^A$, $A=r,a$ and use for the bulk metric the signature $+-++$. The first direction  $r$ is the holographic one and we will use $a,b,c,\ldots =0,1,2$ for transverse Lorentz indices along with $\alpha,\beta,\gamma=1,2$. Coordinate indices will be denoted $\mu,\nu,\rho, \ldots$ and $i,j,k, \ldots$ for transverse spacetime and spatial directions respectively, with $\mathrm{x}\equiv(t,x^1,x^2)\equiv(t,x)$.

Bulk solutions are taken in the Fefferman--Graham (FG) form  
\beq
\label{FGform}
\mathrm{d}s^2 = \frac{L^2}{r^2}\mathrm{d}r^2 +\frac{r^2}{L^2}\eta_{ab}E^a(r,\mathrm{x})E^b(r,\mathrm{x})\,.
\eeq
For torsionless connections there is always a suitable gauge choice 
such that the metrics (\ref{FGform}) are fully determined by two coefficients $\hat{e}^a$ and $\hat{f}^a$ in the 
expansion of the coframe one-forms $\hat{E}^a(r,\mathrm{x})$ along the holographic coordinate $r\in [0,\infty)$
\beq
\label{vielbeil}
\hat {E}^a(r,\mathrm{x})= \left[\hat{e}^a(\mathrm{x})+\frac{ L^2}{r^2} \hat{F}^a(\mathrm{x})+\cdots\right]+\frac{ L^3}{r^3}\left[\hat{f}^a(\mathrm{x})+\cdots \right]\,.
\eeq
The asymptotic boundary is at   $r\to\infty$. The ellipses in (\ref{vielbeil}) denote terms that are multiplied by higher negative powers of $r$. 
Their coefficients 
are determined by $\hat{e}^a$ and $\hat{f}^a$, and have specific geometrical interpretations\footnote{For example, the coefficient $\hat{F}^a$ is related to the boundary Schouten tensor.}, though this is not relevant for our discussion.

The $3+1$-split formalism makes clear that $\hat{e}^a(\mathrm{x})$ and $\hat{f}^a(\mathrm{x})$, being themselves vector-valued one-forms in the boundary, are the proper  canonical variables playing the role of boundary ``coordinate'' and ``momentum'' for the (hyperbolic) Hamiltonian evolution along $r$. For the stationary backgrounds under consideration, describing thermally equilibrated non-dissipating boundary fluid configurations, $\hat{e}^a$ and $\hat{f}^a$ are $t$-independent.   

The boundary ``coordinate'' is given by the set of one-forms $\hat{e}^a$. The corresponding dual vector fields are 
\be
\check{e}_a\,,\quad\hat{e}^a(\check{e}_b)=\delta^a_b\,.
\ee
 These provide the boundary orthonormal frame with  metric given by the symmetric $(0,2)$-tensor 
 \be
 \label{orthometRan}
 \hat g=\eta_{ab}\hat{e}^a\otimes\hat{e}^b\,.
 \ee
 For this coframe we must  determine the ``momentum'' of the boundary data. For example, when the boundary data carry zero mass, we expect this to be zero. In this case $\hat{f}^a(x)=0$ and the unique exact solution of the Einstein's equations is pure AdS$_4$. 

More generally, the vector-valued one-form $\hat{f}^a$ satisfies 
\beq
\label{condfa}
\hat{f}^a\wedge \hat{e}_a=0\,,\quad\epsilon_{abc}\hat{f}^a\wedge \hat{e}^b\wedge \hat{e}^c=0\,,\quad \epsilon_{abc}\mathrm{D}\hat{f}^b\wedge \hat{e}^c=0\,,
\eeq
where the action of the generalized exterior derivative $\mathrm{D}$ on a vector-valued one-form $\hat{V}^a$ is defined as
\beq
\label{covD}
\mathrm{D}\hat{V}^a=\mathrm{d}\hat{V}^a +\epsilon^{a}_{\hphantom{a}bc}\hat{B}^b\wedge\hat{e}^c\,,
\eeq
and the ``magnetic field'' $\hat{B}^a$ is the Levi--Civita spin connection associated with $\hat{e}^a$ \cite{Mansi:2008br}.
One can easily see that the conditions (\ref{condfa}) imply, respectively, the symmetry, tracelessness and covariant conservation of the $(1,1)$-tensor $T=T^a_{\hphantom{a}b}\check{e}_a\otimes \hat{e}^b$, defined as 
\beq
\label{fT}
\hat{f}^a=\frac{1}{\kappa}T(\hat{e}^a)=\frac{1}{\kappa}T^a_{\hphantom{a}b}\hat{e}^b
\,,\quad \kappa=\frac{3}{8\pi G_\mathrm{N} L}\, .
\eeq
Hence we can interpret the latter as the covariantly conserved energy--momentum tensor of a conformal field theory. 

\subsection{The energy--momentum tensor of holographic fluids}\label{emhol}
The next step is to set the relationship of the one-form (\ref{fT}) with a {\it relativistic fluid}. This requires the identification of a timelike and normalized velocity field $\check{u}$ and its corresponding one-form $\hat{u}$ 
\beq
\check{u}=u^a\check{e}_a\, ,\quad\hat{u}=u_a\hat{e}^a\,,\quad\eta_{ab}u^au^b=u^au_a=-1
\eeq
with respect to the boundary frame. From the outset, the boundary frame does not have to be comoving.  The vector-valued one-form $\hat{f}^a$ can be decomposed in longitudinal and transverse components with respect to the fluid velocity $\check{u}$. The decomposition is performed by introducing the longitudinal and transverse projectors:
\begin{equation}
\label{proj}
U^a_{\hphantom{a}b} = - u^a u_b\, , \quad h^a_{\hphantom{a}b} =  u^a u_b + \delta^a_{b}\,.
\end{equation}
This allows to express the coframe components as $
\hat{e}^a = \left(U^a_{\hphantom{a}b}+ h^a_{\hphantom{a}b} \right)\hat{e}^b
$.
Consequently the vector-valued ``momentum'' form reads in general:
\beq
\label{bmomentum}
\hat{f}^a = {\cal E}U^a_{\hphantom{a}b}\hat{e}^b +{\cal P} h^a_{\hphantom{a}b}\hat{e}^b= ({\cal E}+{\cal P})u^a\hat{u} +{\cal P}\hat{e}^a
\,,
\eeq
where $\mathcal{E}$ and $\mathcal{P}$ are \emph{a priori} $x$-dependent coefficients. Holographic hydrodynamics then advocates the interpretation of (\ref{bmomentum}) as the energy--momentum tensor of a relativistic fluid that satisfies the laws of thermodynamics. In particular, all dissipative and non-dissipative terms in a suitable derivative expansion of the velocity field can be classified and related to transport coefficients \cite{Kovtun,Minwalla}. Here we are interested in 
particular stationary bulk solutions for which we expect the energy--momentum tensor be reduced to the perfect relativistic form
\be
\label{fluidemST}
T^a_{\hphantom{a}b} = (\varepsilon +p) u^au_b +p\delta^a_{b}\,.
\ee
 Inserting (\ref{bmomentum}) in (\ref{fT}) one recovers indeed (\ref{fluidemST}) with $\varepsilon=\kappa{\cal E}$ and $p=\kappa{\cal P}$, where $\varepsilon$ and $p$ are the energy and pressure densities measured in the comoving frame. Tracelessness implies $\varepsilon=2p$, {\it i.e.,} the fluid is conformal. This structure (\ref{fluidemST}) will emerge explicitly in the KAdS and TNAdS backgrounds, and the background velocity field will be shearless, expansionless and non-accelerating, but will generically have vorticity. 

Several remarks are in order at this stage, concerning the absence of dissipative phenomena in a fluid, which is not a priori inviscid. This can only occur if the fluid has a specific kinematic configuration, making its viscosity ignorable. The first derivative of a vector-field congruence $\check{u}$
is captured by its acceleration, shear, expansion and vorticity\footnote{Expressions (\ref{def1}), (\ref{def22}) and (\ref{def23}) are valid only in 3 spacetime dimensions.}: 
\begin{equation}
\label{def1}
\nabla_{a} u_b=-u_a a_b +\sigma_{ab} +\frac{1}{2}\Theta h_{ab}+\omega_{ab}
\end{equation}
with
\begin{eqnarray}
a_a&=&u^b\nabla_bu_a\label{def21}\\
\sigma_{ab}&=&\frac{1}{2} h_a^{\hphantom{a}c} h_b^{\hphantom{b}d}\left(
\nabla_c u_d+\nabla_d u_c
\right)-\frac{1}{2} h_{ab}h^{cd} \nabla_c u_d \label{def22}\\
\Theta&=&\nabla_a u^a
\label{def23}\\
\omega_{ab}&=&\frac{1}{2} h_a^{\hphantom{a}c} h_b^{\hphantom{b}d}\left(
\nabla_c u_d-\nabla_d u_c
\right)\,,\label{def24}
\end{eqnarray}
the latter being also expressible as a 2-form
\begin{equation}\label{def3}
\omega= \frac{1}{2}\omega_{ab} \hat{e}^a\wedge\hat{e}^b =\frac{1}{2}\left(\mathrm{d}\hat{u} +
\hat{u} \wedge\hat{a}\right)\, .
\end{equation}
These tensors satisfy several simple identities:
\begin{equation}
u^a \sigma_{ab}=0\,,\quad u^a \omega_{ab}=0\,, \quad u^a a_a=0\,, \quad u^a \nabla_b u_a=0\,, \quad h^c_{\hphantom{c}a} \nabla_b u_c =\nabla_b u_a\,  .
\end{equation}
As mentioned above, the exact solutions that we will study in this paper give rise to fluids with $a_a =\sigma_{ab}=\Theta=0$ and $\omega\neq 0$. In a nutshell, this is due to the following fact: in the generic stationary backgrounds under consideration, the velocity field is a constant-norm, timelike Killing congruence. Such congruences are always geodesic, shearless and expansionless. 

Transport properties, such as shear viscosity, bulk viscosity and (dissipationless) Hall viscosity, can be studied by perturbing the fluid about the above stationary, shearless, expansionless and non-accelerating state. The most general form for the viscous stress current at leading order in fluctuations is
reads: 
\begin{equation}
\label{Tvisc}
\kappa \hat f^a_{\mathrm{visc.}}=-2\eta \sigma^{a}_{\hphantom{(a}b}\hat{e}^b-\zeta \Theta h^{a}_{\hphantom{a}b}\hat{e}^b
+\zeta_{\mathrm{H}} \epsilon^{(a}_{\hphantom{a}bc} u^b\sigma^{cd)}\eta_{de}^{\vphantom{b}}\hat e^e\, ,
\end{equation}
where $\eta$ is the  shear viscosity,  $\zeta$ the bulk viscosity and  $\zeta_{\mathrm{H}}$ the Hall viscosity.  In the case of a conformal fluid, the equation of state is $\varepsilon=2p$, whereas the product $\zeta \Theta$ must vanish for any kinematical configuration, which implies that $\zeta = 0$.
The background value of $f^a_{\mathrm{visc.}}$ vanishes provided the velocity field has neither shear nor expansion. As we will see in the forthcoming sections, this happens indeed for the general class of boundary metrics of the Papapetrou--Randers form (\ref{Papa}), with \emph{fluids at rest in the Papapetrou--Randers frame} ($\check{u}=\check{e}_0$). Let us finally mention that for a shearless 
conformal fluid, the conservation of the energy--momentum tensor  leads to the following equations of motion (see e.g. \cite{Caldarelli:2008mv, Caldarelli:2008ze}):
\begin{equation}
\nabla_{\check{u}}p=0\, , \quad \hat{a}=-\frac{\nabla_{\perp}p}{3p}\label{Tcons}
\end{equation}
($\nabla_{\perp}=\nabla+\hat{u}\nabla_{\check{u}}$ stands for the covariant derivative along the direction normal to the velocity field $\check{u}$). The first equation requires $p$ to be constant along the fluid lines. From the second we learn that if the fluids are inertial, $p$ is also constant in the normal surfaces, \emph{i.e.} strictly constant everywhere. Again, this requirement will be met in the cases of study here.

Although the two necessary ingredients for the description of a relativistic perfect fluid, namely the boundary frame and the velocity one-form, are nicely packaged in the leading and subleading independent boundary data, 
until now we did not assume any specific relationship between them.  Nevertheless it is clear that such a relationship would be imposed by any {\it exact} solution of the bulk gravitational equations, given the interior boundary conditions. We will soon observe that the FG expansion of the exact solutions Schwarzschild--AdS$_4$,  Kerr--AdS$_4$  and Taub--NUT--AdS$_4$ yield the {\it same form} for the boundary energy--momentum tensor, namely
\beq
\label{efrel}
\hat{f}^0=-\frac{2M}{3L}\hat{e}^0\,,\quad \hat{f}^\alpha=\frac{M}{3L}\hat{e}^\alpha\,.
\eeq
The boundary frame one-forms $\hat{e}^a$ are themselves, of course, different in the three solutions. Comparing (\ref{fT}), (\ref{fluidemST}) and (\ref{efrel}), we find 
\beq\label{conpress}
\varepsilon = 2p=2\kappa\frac{M}{3L}\,,
\eeq
constant as already advertised.
The above three exact solutions describe thus the  \emph{same conformal fluid} in different kinematical states. More importantly, (\ref{efrel}) fixes the direction of the velocity field with respect to the boundary frame to be 
\beq
\label{comframe}
\check{u}=\check{e}_0\,.
\eeq
In the Papapetrou--Randers geometry (see below, Eq. (\ref{Papa})), this congruence is tangent to a constant-norm Killing field and has thus zero shear, expansion and acceleration  (consistent, according to \eqref{Tcons}, with the constant pressure found in \eqref{conpress}). 
It also shows that the observer's frame $\check{e}_a$ is {\it comoving}.  Therefore, in the FG expansion the kinematic properties of holographic fluids are determined by the geometric properties of the boundary  comoving frame. 

\subsection{Papapetrou--Randers, Zermelo and relationship with analogue gravity}\label{aether}

It should be noted that in a holographic setup the boundary spacetime is completely filled with our stationary holographic fluid. Consequently the latter acts as a medium (\emph{\ae ther}) for both sound and light propagation. In particular, the velocity of light in the boundary of holographic hydrodynamics does not coincide with the velocity of light in the vacuum.  This issue has not arisen so far since light propagation in the boundary has not been studied in holographic hydrodynamics. It is connected to the notorious problem of making the boundary photon dynamical\footnote{Other optical properties of the boundary fluid have been studied in \cite{Policastro}.}. This said, it is still instructive to 
study  geometrical optics in the boundary, ignoring the fluid and its holographic origin. As we will see, it allows to set connections with some quite different physical setups.

Light travels in null geodesics of the boundary metric. This is a trivial problem when the latter is Minkowski, but becomes interesting when rotation is involved. The boundary metrics for the KAdS and TNAdS bulk geometries have the following general  stationary form, which will be referred to as Papapetrou--Randers\footnote{This form was used by Papapetrou \cite{Papapetrou} to find exact axisymmetric solutions of Einstein's equations. It was also termed the Randers form of a stationary metric in \cite{Gibbons}. In our case, however, the metrics (\ref{Papa}) are {\it  not} solutions of the three-dimensional vacuum Einstein's equations -- they can be solutions in the presence of appropriate sources. Note also that Eq. (\ref{Papa}) is a specific representative of the conformal class of stationary boundary metrics; this choice has the property of being non-accelerating.} (PR):
\beq
\label{Papa}
\mathrm{d}s^2=-(\mathrm{d}t-b_i(x) \mathrm{d}x^i)^2+a_{ij}(x)\mathrm{d}x^i \mathrm{d}x^j\,.
\eeq
General geodesics in PR geometry can be shown to satisfy the following equations \cite{Gibbons} (obtained via a Kaluza--Klein-like procedure, as expected from the form of (\ref{Papa})): 
\beq 
\label{magnflow}
\frac{\mathrm{D}v^i}{\mathrm{d}\ell} = \frac{1}{\sqrt{2{\mathcal{E}}}}  F^i_{\hphantom{i}j} v^j, \quad v^j=\frac{\mathrm{d}x^j}{\mathrm{d}\ell}. 
\eeq
The latter expression is two-dimensional and is meant to provide the geometric locus of the spatial tracks of the geodesics (as opposed to the full spacetime orbits). Those are parameterized in terms of 
 $\ell$, the arc length in the spatial metric:
\begin{equation}\label{ransp}
\mathrm{d}\ell^2= a_{ij}
\mathrm{d}x^i\mathrm{d}x^j\,,
\end{equation}
and $\nicefrac{\mathrm{D}}{\mathrm{d}\ell}$ is the corresponding Levi--Civita covariant derivative. 
The constant $\mathcal{E}$ is the conserved ``energy'' associated with the Killing vector $\partial_t$. 
In Eq. (\ref{magnflow}), we introduced 
\begin{equation}\label{mf}
F_{ij}=\partial_i b_j-\partial_jb_i\,,
\end{equation}
which will be later on identified with the vorticity of the fluid, and appears here as a ``magnetic field'' in the two-dimensional space. 
Notice also that the \emph{null} geodesics of (\ref{Papa}), satisfying (\ref{magnflow})  with $\mathcal{E}=\nicefrac{1}{2}$ ($\mathcal{E}$ is smaller or larger than this value, for timelike or spacelike geodesics, respectively), are also geodesics of the following asymmetric Finslerian norm introduced by Randers:
\beq
\label{Randers}
{\cal F}(x,v) =\sqrt{a_{ij}v^i v^j}+b_i v^i\,, \quad v^i\in {\bf T}_x{\cal M}\,.
\eeq

Particles in magnetic fields undergo cyclotron motion, known to be similar to inertial motion observed from a rotating frame. The above magnetic analogue of geodesic motion in PR geometries (valid actually in any dimension) provides therefore an indirect, though very physical, hint on how an observer perceives the fluid's rotation -- which turns out to be inertial for the holographic fluids, as we will shortly see. It also reveals the risk of existence of CTCs, indeed inherent in geometries of the type (\ref{Papa}).

Even though geodesic congruences are  intrinsic data of a geometry, their perception depends on the frame of reference. The above magnetic analogue carried out in Eqs. (\ref{magnflow}) was reached by choosing a specific frame $\check{e}_a$ evolving in time along $\partial_t\equiv \check{e}_0$ with 
\begin{equation}
\label{coframeRanders}
\hat{e}^0 =  \mathrm{d}t-b ,\quad  \hat{e}^\alpha=E^\alpha_{\hphantom{\alpha}i}\mathrm{d}x^i, \quad E^\alpha_{\hphantom{\alpha}i}E^\beta_{\hphantom{\beta}j}\delta_{\alpha\beta}=a_{ij},
\end{equation}
so as (\ref{Papa}) be of the orthonormal form (\ref{orthometRan}).
This will be referred to as the Papapetrou--Randers frame, in which the inertial motion appears as cyclotron motion. Alternative frames exist, where the rotation is intrinsically or partly attributed to the motion itself. We will quote and use a particular one, appealing for the bridge it sets with analogue gravity set ups: the Zermelo frame.

Consider a two-dimensional Euclidean manifold ${\cal M}$ with metric $h_{ij}$. The so-called Zermelo navigation problem \cite{Zer31} asks for the minimum-time trajectories on that manifold under the influence of a time-independent wind $W^i$. Remarkably, it was shown in \cite{Shen} that these coincide {\it exactly} with the null geodesics of the Randers form (\ref{Randers})  provided the Randers data $(a_{ij},b_i)$ are related to the Zermelo data $(h_{ij},W^i)$ as
\bea
\label{RZ1}
&& a_{ij}= \frac{h_{ij}}{\lambda}+ \frac{W_i W_j}{\lambda^2}\,,\quad\lambda = 1-h_{ij}W^i W^j\,,\\
\label{RZ2}
&&b_i=-\frac{W_i}{\lambda}\,,\quad W_i=h_{ij}W^j\,,\quad a^{ij}=\lambda(h^{ij}-W^i W^j)\,.
\eea
Using the above, the metric (\ref{Papa}) takes the form
\beq
\label{Zermelo}
\mathrm{d}s^2=\frac{1}{\lambda(x)}\left[-\mathrm{d}t^2+h_{ij}(x)\left(\mathrm{d}x^i-W^i(x) \mathrm{d}t\right)\left(\mathrm{d}x^j-W^j(x) \mathrm{d}t\right)\right]\,,
\eeq
which is called the Zermelo form of the stationary geometry (\ref{Papa}). 

The timelike component of the corresponding coframe is denoted as $\hat{z}^0=\nicefrac{\mathrm{d}t}{\sqrt{\lambda}}$ (vs. $\hat{e}^0=\mathrm{d}t-b$ in PR) and the spacelike ones as $\hat{z}^\a$. One can also define (see Sec. \ref{KTNfl}) the set of dual frames $\check{e}_a$ and $\check{z}_a$.
Being both orthonormal, they are related by a local Lorentz transfromation:
\beq
\label{P-RZtransf}
\hat{z}^a=\Lambda^a_{\hphantom{a}b}\hat{e}^b
\eeq
with the explicit form of the matrix elements $\Lambda^a_{\hphantom{a}b}$ given in (\ref{LambdaZerm}). Hence, the stress current for the Zermelo observer is
\beq
\label{fzerm}
\hat{f}^a_{\mathrm{Z}}=\Lambda^a_{\hphantom{a}b}\hat{f}^b = \frac{\varepsilon +p}{\kappa}\hat{u}u^a_{\mathrm{Z}}+ \frac{p}{\kappa}\hat{z}^a\,.
\eeq
It describes a fluid moving with a velocity field given by (see also (\ref{uzerm2}) below)
\beq
\label{uzerm}
\check{u}=u^a\check{z}_a\,,\quad u^a_{\mathrm{Z}}=\Lambda^a_{\hphantom{a}b}u^b=\Lambda^a_{\hphantom{a}0}=
\left(\begin{smallmatrix}
      \nicefrac{1}{\sqrt{\lambda}} \\ - \nicefrac{W^\a}{\sqrt{\lambda}}
    \end{smallmatrix}\right)
\,,\quad W^\a=\sqrt{\lambda}L^\a_{\hphantom{\alpha}i}W^i\,,
\eeq
where $L^a_{\hphantom{\alpha}i}$ are defined in (\ref{coframe_Zermelo}). In other words, the Zermelo frame description is that of a fluid boosted with velocity $W^\a$ and Lorentz factor $\nicefrac{1}{\sqrt{\lambda}}$. 

We finally come to the advertised relationship with analogue gravity systems, directly set in the Zermelo frame (\ref{Zermelo}). Metrics of that form are named acoustic or optical and are used for describing the propagation of sound/light disturbances in relativistic or non-relativistic fluids moving with velocity $W^i$ in the spatial  geometry $h_{ij}$, and subject to appropriate thermodynamic/hydrodynamic assumptions (e.g. barotropicity for sound propagation). More precisely, sound and light propagation in analogue gravity systems is effectively equivalent to massless fields propagating in the Zermelo geometry (\ref{Zermelo}) \cite{Barcelo}. We can therefore consider that the sound/light fluctuations in a moving medium define a set of {\it Zermelo observers}. In this approach, the metric (\ref{Zermelo}) appears as analogue metric and is \emph{not} the actual metric of physical spacetime\footnote{In particular, when the fluid is non-relativistic, there is no physical spacetime metric. but there is still an analogue metric  (\ref{Zermelo})} . Under this perspective, peculiarities such as CTCs, potentially present in the analogue geometry, have no real, physical existence. They are manifestations of other underlying physical properties such as supersonic/superluminal regimes in the flowing medium. 

What we have shown here is that holography provides a natural construction of a set of Zermelo observers {\it plus} an underlying fluid with concrete thermodynamic properties. This is an explicit setup where the propagation of various modes can be studied and hence, interesting properties of analogue gravity systems could be uncovered. 

To conclude, the holographic hydrodynamic regimes under consideration can support two alternative interpretations: the direct description of a conformal fluid moving in a genuine spacetime of the PR or Zermelo forms (\ref{Papa}) and (\ref{Zermelo}) (the latter differ in the choice of the frame, but describe the same spacetime geometry), or, the indirect description of sound/light propagation in a flowing fluid with velocity $W^i$ in a space geometry $h_{ij}$. In this case
holography gives a handle on the physics of the relevant sound/light modes propagating in the moving medium.

\section{Kerr--AdS$_4$ and Taub--NUT--AdS$_4$ geometries}\label{geom}

Our two explicit examples of holographic fluids with vorticity reside on the boundary of the Kerr--AdS$_4$ (KAdS) and Taub--NUT--AdS$_4$ (TNAdS) exact solutions of vacuum Einstein equations with negative cosmological constant. We recall here some of the salient features of these geometries and in particular of the (Lorentzian) TNAdS which seems to be less well known. 

\subsection{Kerr--AdS$_4$}

The four-dimensional Kerr solution of Einstein's equation with cosmological constant $\Lambda=-3k^2=-\nicefrac{3}{L^2}$ reads: 
\begin{eqnarray}
 \mathrm{d}s^2 &=& \frac{\mathrm{d}r^2}{V(r, \theta)}  -V(r, \theta)
      \left[ \mathrm{d}t - \frac{a}{\Xi} \sin^2\theta  \mathrm{d}\phi\right]^2\nonumber\\
  &&+ \frac{\rho^2}{\Delta_\theta} \mathrm{d}\theta^2
  + \frac{\sin^2\theta \Delta_\theta}{\rho^2}\left[a \mathrm{d}t - 
      \frac{r^2+a^2}{\Xi}  \mathrm{d}\phi \right]^2,
      \label{KAdS}
\end{eqnarray}
where
\begin{equation}
V(r,\theta)=\frac{ \Delta_r }{\rho^2}
\end{equation}
and
\begin{eqnarray}
 \Delta_r & = & (r^2 + a^2)(1+ k^2 r^2) - 2Mr \\
  \rho^2 & = & r^2 + a^2\cos^2\theta \\
 \Delta_\theta & = & 1 - k^2 a^2 \cos^2\theta  \\
 \Xi & = & 1 - k^2 a^2.\label{KXi}
\end{eqnarray}
The solution at hand describes the field generated by a rotating mass.
The ADM mass and angular momentum of the black hole can be computed using the Hamiltonian approach \cite{Henneaux:1985tv}: 
\begin{equation}
\label{ADMmJ}
m=\frac{M}{\Xi^2},\quad 
J=\frac{aM}{\Xi^2}.
\end{equation}
The geometry has inner ($r_-$) and outer ($r_+$) horizons, where  $ \Delta_r$ vanishes,  as well as an ergosphere at $g_{tt}=0$. One can show that the rotating AdS black hole is stable for $a^2<k^2$ \cite{Hawking:1998kw,Awad:1999xx,Caldarelli:1999xj, Gibbons:2004ai}, hence the asymptotically flat black hole ($k=0$) is unstable. This is a consequence of frame dragging
(behind the ergosphere no static observer exists), which disappears asymptotically in the Kerr black hole, but persists in the Kerr--AdS. 

On the outer horizon $\Delta_r(r_+)=0$, any fixed-$\theta$ observer has a determined angular velocity:
\begin{equation}
\Omega_{\mathrm{H}}=\frac{a\Xi}{r^2_++a^2},
\end{equation}
and thus a tangent vector proportional to 
\begin{equation}
\label{Kilbolt}
\partial_t+\Omega_{\mathrm{H}}\partial_\phi,
\end{equation}
which is light-like.
One should stress here that the angular velocity $\Omega_{\mathrm{H}}$ \emph{is not} the one measured at infinity by a \emph{static} observer -- contrary to what happens for the asymptotically flat plain Kerr geometry. In fact, $\Omega_{\mathrm{H}}$ is the angular velocity observed by an asymptotic observer in a natural frame of the coordinate system at hand. This observer is not static, but has an angular velocity
\begin{equation}
\Omega_{\infty} = a k^2,
\end{equation}
which obviously vanishes when the cosmological constant is switched off ($k\to 0$). The angular velocity of the black hole for a static observer at infinity is thus \cite{Gibbons:2004ai}
\begin{equation}
\label{Kerr-ang-vel}
\Omega=\Omega_{\mathrm{H}} +  \Omega_{\infty} = \frac{a(1+ k^2 r_+^2)}{r^2_++a^2}.
\end{equation}

When moving to the Euclidean ($t=-i\tau, a=i\alpha $), for analyzing e.g. the thermodynamical properties of the background,  the outer horizon appears as a bolt \emph{i.e.} as a two-dimensional fixed locus of the Killing vector (\ref{Kilbolt}) -- in its Euclidean version. On this bolt\footnote{The general classification of fixed points was originally presented in \cite{Gibbons:1979xm}.}, the northern and southern poles appear as extra zero-dimensional fixed points (nuts) of another Killing vector, $\partial_\phi$. This has led several authors to interpret the Kerr black hole as a nut--anti-nut bound state connected by a Misner string \cite{Hunter:1998qe,Manko:2009xx}.

Anticipating the analysis of the forthcoming sections, we obtain for the Kerr metric (\ref{KAdS})
\begin{equation}
\label{KAdS-bry}
\mathrm{d}s^2_{\mathrm{bry.}}=\lim_{r\to \infty}\frac{\mathrm{d}s^2}{k^2r^2}=  -
      \left[ \mathrm{d}t - \frac{a}{\Xi} \sin^2\theta  \mathrm{d}\phi\right]^2
      + \frac{\mathrm{d}\theta^2}{k^2\Delta_\theta}
  +  \frac{\Delta_\theta}{k^2\Xi^2} \sin^2\theta \mathrm{d}\phi^2.
\end{equation}
This boundary metric can be recast in several ways:
\begin{eqnarray}
\mathrm{d}s^2_{\mathrm{bry.}}&=& \frac{\Delta_\theta}{\Xi}\left(-\mathrm{d}t^2+ \frac{\Xi}{k^2\Delta_\theta^2}\left(\mathrm{d}\theta^2+  \frac{\Delta_\theta}{\Xi}  \sin^2\theta \left[ \mathrm{d}\phi + \Omega_{\infty} \mathrm{d}t \right]^2 \right)\right)
 \label{KAdS-zer}\\
 &=& \frac{1}{\Delta_{\theta'}}\left(-\mathrm{d}t^2+ \frac{1}{k^2}\left(\mathrm{d}\theta'^2+   \sin^2\theta' \left[ \mathrm{d}\phi + \Omega_{\infty} \mathrm{d}t \right]^2 \right)\right).
\label{KAdS-einrot}
\end{eqnarray}
The last expression is obtained by trading $\theta$ for $\theta'$ as
\begin{equation}
\Delta_{\theta}  \Delta_{\theta'} =\Xi,
\end{equation}
where 
$ \Delta_{\theta'} =  1 - k^2 a^2 \cos^2\theta'$. It describes the boundary of Kerr--AdS as conformal to the three-dimensional Einstein universe, rotating at angular velocity $\Omega_{\infty}$
\cite{Hawking:1998kw}.

A last feature we wish to mention in relation with the boundary geometry of Kerr--AdS is the behavior around the poles, at $\theta \approx 0 \ \mathrm{or} \ \pi $:
\begin{equation}
\label{KAdS-bry-pol}
\mathrm{d}s^2_{\mathrm{bry.}}\approx-\left(
\mathrm{d}t - \Omega_{\infty}\chi^2\mathrm{d}\phi
\right)^2 +
\mathrm{d}\chi^2
+\chi^2
\mathrm{d}\phi^2
\end{equation}
with $\chi=\frac{\theta}{k\sqrt{\Xi}}$ for the northern pole, and  $\chi=\frac{\pi-\theta}{k\sqrt{\Xi}}$ for the southern pole. Metric (\ref{KAdS-bry-pol}) is the Som--Raychaudhuri space, 
found in \cite{SR68} and solving Einstein equations with rotating, charged dust with zero Lorentz force. It belongs to the general family of three-dimensional homogeneous spaces possessing 4 isometries studied in \cite{rayPRD80, rebPRD83}, which include in particular G\"odel space as well as the boundary of the AdS--Taub--NUT space that we will present in the next section. In the case of Som--Raychaudhuri (Eq. (\ref{KAdS-bry-pol})) the isometries are generated by the following Killing vectors:
\begin{equation}
  \begin{cases}
  \label{LKilRn}
K_x=  \frac{\sin\phi}{\chi} \, \partial_\phi-\cos \phi\, \partial_\chi- \Omega_{\infty}\chi\sin \phi  \, \partial_t \\
K_y=  \frac{\cos\phi}{\chi} \, \partial_\phi+\sin \phi \,\partial_\chi- \Omega_{\infty}\chi\cos \phi\, \partial_t  \\
K_0=2 \Omega_{\infty}\, \partial_t\\
K=\partial_\phi.
\end{cases}
\end{equation}
The vectors $K_x, K_y$ and $K_0$ form a Heisenberg algebra, and indeed the Som--Raychaudhuri  metric can be built as the group manifold of the Heiseberg group (Bianchi II) at an extended-symmetry (isotropy) point with an extra symmetry generator\footnote{Recall $\left[K_x, K_y\right]=K_0$, $\left[K_x, K_0\right]=\left[K_y, K_0\right]=0$ and $\left[K, K_x\right]=K_y$ and $\left[K, K_y\right]=-K_x$.} $\partial_\phi$. 

Similarly to G\"odel space, Som--Raychaudhuri  space contains non-geodesic closed time-like curves. These are circles of radius $\chi$ larger than $\nicefrac{1}{\Omega_{\infty}}$ \cite{rebPLA87}. Notice, however, that the boundary of Kerr--AdS is free of closed time-like curves since it is identified with 
Som--Raychaudhuri in a region where $\chi \ll\nicefrac{1}{\Omega_{\infty}}$. As we will see soon, this no longer holds for the boundary of AdS--Taub--NUT.

\subsection{Taub--NUT--AdS$_4$}\label{TNADS}

The Taub--NUT--AdS$_4$ geometry is a foliation over squashed three-spheres solving Einstein's equations:
\begin{eqnarray}
\mathrm{d}s^2 &=&
 \frac{\mathrm{d}r^2}{V(r)}
+\left(r^2+n^2\right)\left((\sigma^1)^2+
\left(\sigma^2\right)^2\right)-4n^2V(r)\left(\sigma^3\right)^2
\nonumber
\label{bulmet}
\\
&=&  \frac{\mathrm{d}r^2}{V(r)}
+\left(r^2+n^2\right)\left(
 \mathrm{d}\theta^2 +
\sin^2\theta \mathrm{d}\phi^2
\right)-4n^2V(r)\left( \mathrm{d}\psi + \cos \theta  \mathrm{d}\phi\right)^2
\label{hyptaubnut-lor}
\end{eqnarray}
with 
\begin{equation}
\label{hyptaubnutpot-lor}
V(r)= \frac{1}{r^2+n^2}\left[
r^2-n^2 -2 M r +k^2\left(
r^4+6n^2r^2 -3 n^4
\right)
\right],
\end{equation}
where 
 \begin{equation}
  \begin{cases}
  \label{LMC}
\sigma^1= \sin\theta \sin\psi \, \mathrm{d}\phi+\cos \psi \, \mathrm{d}\theta \\
\sigma^2= \sin\theta\cos\psi\, \mathrm{d}\phi-\sin\psi\, \mathrm{d}\theta\\
\sigma^3=\cos\theta\, \mathrm{d}\phi+\mathrm{d}\psi
\end{cases}
\end{equation}
are the $SU(2)$ left-invariant Maurer--Cartan forms in terms of Euler angles $0\leq\theta\leq \pi, 0\leq\phi\leq 2\pi, 0\leq\psi\leq {4\pi}$. Besides the mass $M$ and the cosmological constant $\Lambda=-3k^2$, this solution depends on an extra parameter $n$: the nut charge. It is convenient to trade $\psi$ for  $t=-2n (\psi+\phi)$. With this coordinate the metric (\ref{hyptaubnut-lor}) assumes the form
\begin{equation}
\label{hyptaubnutpot-lor-t}
\mathrm{d}s^2 =   \frac{\mathrm{d}r^2}{V(r)}
+\left(r^2+n^2\right)\left(
 \mathrm{d}\theta^2 +
\sin^2\theta \mathrm{d}\phi^2
\right)-V(r)\left[\mathrm{d}t + 4n\sin^2 \frac{\theta}{2}  \mathrm{d}\phi\right]^2.
\end{equation}

The original Taub--NUT solution  \cite{Taub:1950ez,NUT} was a vacuum solution designed for cosmology (see also \cite{HE73} for a more detailed analysis). Since then, many variants have been studied, both with Lorentzian and Euclidean signature (reached by setting $\nu=in$ and $\tau =it$), with or without cosmological constant or mass. In the Euclidean version and in the absence of mass, one finds the original Eguchi--Hanson and Taub--NUT gravitational instantons for vanishing cosmological constant and the Fubini--Study solution with cosmological constant (see \cite{Eguchi:1978gw,Eguchi:1979yx, P85}). The former are self-dual and the latter quaternionic (Weyl-self-dual).
Adding a mass opens up new possibilities according to the kind of horizons that appear, and the corresponding solutions can be either (Weyl-)self-dual or not, such as Taub--bolt, Pedersen, etc. Self-duality or quaternionic self-duality must be abandoned in the Lorentzian framework. We will not pursue further this general description (interesting related information can be found in e.g.  \cite{Zoubos:2002cw, PVH12}). We will  focus instead on the properties of the specific Lorentzian metric (\ref{hyptaubnutpot-lor-t}).

In the following, we provide the prominent properties of the geometry (\ref{hyptaubnutpot-lor-t}). Many of these properties are a consequence of its isometry group $SU(2)\times U(1)$, generated by the Killing vectors
  \begin{equation}
\begin{cases}
  \label{LKilR2}
\xi_1= - \sin\phi \cot\theta\, \partial_\phi+\cos \phi\, \partial_\theta-2\nu\frac{\sin \phi}{\sin \theta}(1-\cos \theta) \, \partial_t \\
\xi_2=  \cos\phi \cot\theta\, \partial_\phi+\sin \phi \,\partial_\theta+2\nu\frac{\cos \phi}{\sin \theta}(1-\cos \theta) \, \partial_t  \\
\xi_3= \partial_\phi-2\nu\, \partial_t\\
e_3= -2\nu\, \partial_t.
\end{cases}
\end{equation}
Two extra vectors $e_1$ and $e_2$ generate with $e_3$ the right $SU(2)$. These are not Killing, however, due to the squashing of the spherical leaves.

The solution at hand has generically two horizons ($V(r_\pm)=0$) and is well-defined outside the outer horizon $r_+$, where $V(r)>0$. In the Euclidean language, this horizon is a bolt\footnote{By analytic continuation, the solution (\ref{hyptaubnut-lor}) with (\ref{hyptaubnutpot-lor}) is mapped onto the so-called AdS--Taub-bolt.} \emph{i.e.} the two-dimensional  fixed locus of the Killing vector $e_3$. On this surface, $\theta = \pi$ is an isolated fixed point of another Killing vector $\xi_3+e_3$. This is a nut, carrying a net nut charge $n$. 

The nut is the origin of a Misner string \cite{misner:1963}, departing from $r=r_+$, all the way to $r\to\infty$, on this southern pole at $\theta=\pi$. The geometry is nowhere singular along the Misner string, which appears as a coordinate artifact much like the Dirac string of  a magnetic monopole is a gauge artifact. In order for this string to be invisible, coordinate transformations displacing the string must be univalued everywhere, which is achieved by requiring the periodicity condition $\psi\equiv \psi + 4\pi$ or equivalently $t\equiv t - 8\pi n$. Alternatively, one can avoid periodic time and keep the Misner string as part of the geometry. This semi-infinite spike appears then as a source of angular momentum, integrating to zero \cite{bonnor:1969, dowker:1974}, and movable at wish using the transformations generated by the above vectors. This will be our viewpoint throughout this work.
However, despite the non-compact time, the AdS--Taub--NUT geometry is plagued with closed time-like curves, which disappear only in the vacuum limit $k\to 0$ \cite{Astefanesei:2004kn}. Even though this is usually an unwanted situation, it is not sufficient for rejecting the geometry, which from the holographic perspective has many interesting and novel features, as we will see later.

Finally, we would like to mention the behavior of the AdS--Taub--NUT geometry at large $r$:
\begin{eqnarray}
\mathrm{d}s^2_{\mathrm{bry.}}&=&
\frac{1}{k^2}\left((\sigma^1)^2+
\left(\sigma^2\right)^2-4 k^2n^2\left(\sigma^3\right)^2\right)
\nonumber
\label{bounmet}\\
&=& -\left[\mathrm{d}t + 4n\sin^2 \frac{\theta}{2} \mathrm{d}\phi\right]^2+\frac{1}{k^2} \left(
 \mathrm{d}\theta^2 +
\sin^2\theta \mathrm{d}\phi^2
\right).
\label{anagoed}
\end{eqnarray}
This is a squashed three-sphere appearing as a limiting leave of the foliation (\ref{hyptaubnut-lor}). The squashing is Lorentzian as in the bulk,  and consequently the closed time-like curves survive on the boundary. This space is homogeneous and belongs to the already quoted family of spaces invariant under a four-parameter group of motions \cite{rayPRD80, rebPRD83}, here generated by the vectors (\ref{LKilR2}). 

Zooming around the northern pole exhibits  the Som--Raychaudhuri  metric (\ref{KAdS-bry-pol}), as in the AdS--Kerr, with $\Omega_\infty$  traded for $- n k^2 $ and $\chi=\nicefrac{\theta}{k}$.  This corresponds to a contraction of $SU(2)\times U(1)$ into a semi-direct product of the Heisenberg group with an extra $U(1)$ generated by $K_x=-k\xi_1, K_y=k\xi_2, K_0=k^2 e_3, \xi_3=e_3+\partial_\phi$ (see (\ref{LKilRn})). On the southern pole, which is  the track of the Misner string on the boundary, the behavior is somewhat different: 
  \begin{equation}
\mathrm{d}s^2_{\mathrm{bry.}}\approx -\left(\mathrm{d}t + n\left(4-k^2\chi^2\right) \mathrm{d}\phi\right)^2 + \mathrm{d}\chi^2
+\chi^2
\mathrm{d}\phi^2, 
\label{AKT-south}
\end{equation}
where $\chi=\frac{\pi-\theta}{k}$. The latter is known as a flat vortex geometry, homogeneous and invariant under an $E(2)\times U(1)$ algebra\footnote{What we call  $U(1)$ in (\ref{LKilR2}) or (\ref{LLilRs}) is in fact $\mathbb{R}$,  since we eventually consider non-compact $t$.} (Bianchi $\mathrm{VII}_0$) generated by 
  \begin{equation}
\begin{cases}
  \label{LLilRs}
L_x=k \xi_1=  \frac{\sin\phi}{\chi} \left(\partial_\phi -4n\, \partial_t\right)-\cos \phi\, \partial_\chi \\
L_y=-k \xi_2=  \frac{\cos\phi}{\chi} \left(\partial_\phi -4n\, \partial_t\right)+\sin \phi \,\partial_\chi\\
L_0=\xi_3= \partial_\phi -2n\, \partial_t\\
e_3= -2n\, \partial_t.
\end{cases}
\end{equation}

\section{The Kerr and Taub--NUT fluids}\label{KTNfl}

This section is the core of our findings based on two exact Einstein spaces: Kerr--AdS and Taub--NUT--AdS. For both we extract the specific holographic data, which fit the general framework developed in Sec. \ref{holflu}. We discuss in detail the kinematics of the fluid in both PR and Zermelo frames as well as the appearance of the CTCs in TNAdS.

\subsection{The Papapetrou--Randers frame}

As explained in Sec. \ref{holflu}, the FG expansion provides, via the leading and subleading terms, the boundary coframe and the boundary energy--momentum current.  For the backgrounds at hand, namely KAdS (\ref{KAdS}) and TNAdS (\ref{hyptaubnutpot-lor-t}), we easily extract the boundary coframes leading to the boundary geometries (\ref{KAdS-bry}) and (\ref{bounmet}). These are of the general form (\ref{Papa}) with
\be
\label{Kerr_Randers1}
b = \frac{a}{\Xi}\sin^2\theta\ \mathrm{d}\phi\, ,\quad 
a_{ij}=L^2\mathrm{diag}\left(\frac{1}{\Delta_\theta},\frac{\Delta_\theta}{\Xi^2}\sin^2\theta\right)\, 
\ee
for KAdS, while for TNAdS we have instead 
\beq
\label{TN_Randers}
b=-2n(1-\cos\theta) \mathrm{d}\phi\,,\quad a_{ij}=L^2\mathrm{diag}(1,\sin^2\theta).
\eeq
Indeed, the PR orthonormal coframes (\ref{coframeRanders}), as they emerge from the FG expansion, are respectively
\be
\label{KAdS-Randers-cof}
\hat{e}^0=\mathrm{d}t-\frac{a}{\Xi}\sin^2\theta\ \mathrm{d}\phi \, , \quad\hat{e}^1=\frac{L}{\sqrt{\Delta_\theta}} \mathrm{d}\theta\, , \quad\hat{e}^2= 
\frac{L\sqrt{\Delta_\theta}}{\Xi}\sin\theta\ \mathrm{d}\phi 
\, 
\ee
and
\be
\label{TNAdS-Randers-cof}
\hat{e}^0=\mathrm{d}t+2n(1-\cos\theta) \mathrm{d}\phi \, , \quad\hat{e}^1=L \mathrm{d}\theta \, , \quad\hat{e}^2= L\sin\theta\ \mathrm{d}\phi \,.
\ee
The dual boundary frames are of the form
\beq
\label{frameRanders}
\check{e}_0 =  \partial_t\, ,\quad  \check{e}_\alpha=E_\alpha^{\hphantom{\alpha}i}\left(b_i\partial_t+\partial_i\right)\,, \quad E_\alpha^{\hphantom{\alpha}i}E^\beta_{\hphantom{\beta}i}=\delta^\beta_\alpha\, .
\eeq
These vectors can be easily worked out for KAdS and TNAdS, where they are  diagonal, using the above formulas.

Similarly, in both cases, the boundary energy--momentum current is of the form (\ref{efrel}) and describes a conformal fluid with constant pressure (\ref{conpress}) and velocity field (\ref{comframe}): $\check{u}=\partial_t$ and $\hat{u} = -\mathrm{d}t+b$. The fluid is therefore \emph{at rest} in the PR frame, and the corresponding observers are thus comoving. 
Furthermore, $\partial_t$ is a Killing vector with constant norm ($-1$). Hence, its integral lines 
are geodesics: 
\be
\check{a}=\nabla_{\partial_t}\partial_t=0\, .
\ee
The fluid and the comoving observers are inertial. For this geodesic congruence, the shear and expansion systematically vanish, and everything goes as announced in Sec. \ref{holflu}.

The observers accompanying the PR frame, whose tangent bundle is spanned by the vectors $\partial_i$, can define the {\em fluid's physical surface} as the set of points which are synchronous events for their time $t$,  since $\mathrm{d}t( \partial_i)=0$. The tangent planes of this  surface are spanned by $\partial_i$ or by any linear combination of them (orthonormal or not).
The parallel transport of the physical surface along $\check{e}_0$ is the physical manifestation of the fluid's flow in the comoving frame. We in fact find that the physical surface is not parallel transported along $\partial_t$, namely\footnote{Without being exhaustive, we give some of the Christoffel symbols, valid for all PR metrics it the coordinate frame, which can be used to demonstrate some of the quoted properties: $\Gamma^\mu_{tt}=0$, $\Gamma^i_{tj}=-\frac12 a^{ik}(\pa_k b_j-\pa_j b_k)=-a^{ik}\omega_{kj}$ and $\Gamma^t_{ti}=\frac12 (\pa_i b_j-\pa_j b_i)a^{jk}b_k=\omega_{ij}a^{jk}b_k$.\label{conne}}
\beq
\label{cov_alpha}
\nabla_{\partial_t}\partial_i = \omega_{ij}a^{jk}\left(\partial_k+b_k\partial_t\right)
\Leftrightarrow
\nabla_{\check{e}_0}\check{e}_\a= \omega_{\a\b}^{\mathrm{PR}}\delta^{\b\gamma}\check{e}_\gamma
\, ,
\eeq
where $\omega_{ij}=\frac{1}{2}\left(\partial_ib_j-\partial_jb_i\right)$ are  the spacetime components of the vorticity form introduced in (\ref{def3}), which 
reduces here to $\frac{1}{2}\mathrm{d}b$ due to the absence of acceleration, and $\omega_{\a\b}^{\mathrm{PR}}=E_\alpha^{\hphantom{\alpha}i}E_\beta^{\hphantom{\beta}j}\omega_{ij}$ its components in the PR frame.
Hence, the inertial observers perceive the fluid's flow as the rotation upon parallel transport of the geodesic congruence tangent to $\partial_t$. 

For the geometries of the PR form the only non-zero components of the vorticity are along the spatial coframe $\hat{e}^\alpha$. We find for KAdS
\beq
\label{omega_Kerr}
\omega_{\mathrm{K}} = \frac{a}{L^2}\cos\theta \, \hat{e}^1 \wedge \hat{e}^2\,,
\eeq
which describes cyclonic flow (e.g. the motion of the atmosphere of a rotating planet) as seen from the comoving frame. 

In the TNAdS case we must be more careful. Noting that the globally defined one-form is $\hat{e}^2$ rather than ${\mathrm{d}}\phi$, we see that the coefficient $b_2$ in (\ref{TN_Randers}) diverges at $\theta=\pi$. This induces a $\delta$-function singularity in the vorticity
\beq
\label{omega_TN}
 \quad\omega_{\mathrm{TN}}=-\frac{n}{L^2} \, \hat{e}^1 \wedge \hat{e}^2-\frac{n}{L^2}\delta_2(\theta-\pi)\,,
\eeq
where the last term denotes a singular two-form with support only at $\theta=\pi$. The normal part of  (\ref{omega_TN}) describes a vortex flow with constant vorticity.  
The $\delta$-function singularity is the boundary remnant of the Misner string \cite{misner:1963}, which extends, in the chosen coordinates radially along the $\theta=\pi$ axis, intersecting the boundary at a (neutral) {\it ``Misner vortex''}. Since we are not interested in compactifying the Lorentzian time coordinate, this string is physical, \cite{bonnor:1969, dowker:1974} and will have important consequences holographically. The $\delta$-function singularity noted above shows up either as a singular contribution to the torsion of a smooth connection, or equivalently, as a singular contribution to the Levi--Civita connection. 

Before proceeding with the study of an alternative frame, we would like to comment on the magnetic paradigm mentioned in Sec. (\ref{aether}) for generic PR geometries (\ref{Papa}).
According to that analysis, the geodesic motion on the boundary spacetime is analogous to the Newtonian motion of charged particles on the two-dimensional space with metric (\ref{ransp}) subject to the magnetic field (\ref{mf}). On the one hand, for KAdS the space (given in (\ref{Kerr_Randers1})) is a squashed two-sphere, whereas the magnetic field (\ref{omega_Kerr}) is that of a magnetic dipole. On the other hand, in the case of TNAdS, the space is a two-sphere (\ref{TN_Randers}) and the magnetic field (\ref{omega_Kerr}) is generated by a Dirac monopole. In this magnetic picture, the Misner string is traded for a Dirac string, which on the two-dimensional boundary is reduced to a single point: the southern pole.

\subsection{The Zermelo frame}
The space spanned by the vectors $\partial_i$ is tangent to the synchronous surface of the inertial observers. These vectors can be traded for a set of orthonormal vectors
$\check{z}_\alpha$ (see below, Eq. \eqref{frame_Zermelo}), whose space
does not coincide with that spanned by the vectors $\check{e}_\alpha$ orthogonal to $\check{e}_0=\partial_t$. It is then natural to ask what the normalized  timelike vector $\check{z}_0$, orthogonal to $\check{z}_\alpha$, is. Such a choice corresponds to $\check{z}_a$ and $\check{e}_a$ being related by a local Lorentz transformation, as already discussed in Sec. \ref{holflu}, Eq. \eqref{P-RZtransf}. The congruences of $\check{z}_0$ would be the worldlines of a different set of, generally non-inertial, observers. For this set of observers the space spanned by $\check{e}_\alpha$ obeys ${\mathrm{d}}t(\check{e}_\alpha)
-b(\check{e}_\alpha)=0$. Since the Fr\"obenius criterion is not fulfilled (${\mathrm{d}} ({\mathrm{d}}t-b)=-2\omega \Leftrightarrow \left[b_i\partial_t+\partial_i, b_j\partial_t+\partial_j\right]= 2\omega_{ij}\partial_t$), it is not possible to define a universal time whose synchronous hypersurfaces, tangent to $\check{e}_\alpha$, would be the fluid physical surfaces simultaneously for all these observers. 
We find for the frame and the dual coframe: 
\bea
\label{frame_Zermelo}
&\check{z}_0=\frac{1}{\gamma}\left(\partial_t+W^i\partial_i\right)\, , \quad  \check{z}_\alpha=L_\alpha^{\hphantom{\alpha}i}\partial_i&\\
\label{coframe_Zermelo}
&\hat{z}^0=\gamma {\mathrm{d}}t \, , \quad \hat{z}^\alpha= L^\alpha_{\hphantom{\alpha}i}({\mathrm{d}}x^i - W^i {\mathrm{d}}t)&
\eea
with 
\be
\label{lorfac}
\gamma^{-2}=1-a^{ij}b_i b_j\,,\quad W^i=-\gamma^2a^{ij}b_j\, ,\quad 
L_\alpha^{\hphantom{\alpha}i}L^\beta_{\hphantom{\beta}i}=\delta^\alpha_\beta\,.
\ee

In the new orthonormal frame, the boundary metric  assumes the form (\ref{Zermelo}) with
\be
\label{h_lambda}
h_{ij}=\lambda(a_{ij}-b_i b_j)=\lambda L^\alpha_{\hphantom{\alpha}i} L^\beta_{\hphantom{\beta}j}\delta_{\alpha\beta}  
\,,\quad \lambda \equiv\nicefrac{1}{\gamma^2}
\, ,
\ee
which is the  Zermelo form of the metric  (\ref{Papa}). For this reason, we will be referring to the frame $\{\check{z}_a\}$ as the \emph{Zermelo frame}. For KAdS we find
\be
h_{ij}=L^2\mathrm{diag}\left(\frac{\Xi}{\Delta_\theta^2}
, \frac{\sin^2\theta}{\Delta_\theta}
\right)\, ,
\ee
whereas for TNAdS
\be
h_{ij}=\mathrm{diag}\left(
L^2-4n^2\tan^2\nicefrac{\theta}{2},
4\tan^2\nicefrac{\theta}{2}\left(L^2 \cos^2 \nicefrac{\theta}{2}-4n^2  \sin^2 \nicefrac{\theta}{2}\right)^2
\right)\,.
\ee
Again, vectors $\check{z}_\alpha$ can be easily worked out for KAdS and TNAdS.

As quoted in \eqref{uzerm}, the fluid's velocity reads:
\beq
\label{uzerm2}
\check{u}=\gamma\left(\check{z}_0-W^\alpha\check{z}_\alpha\right)\,,\quad W^\alpha =\frac{1}{\gamma}L^\alpha_{\hphantom{\alpha}i}W^i\, ,
\eeq
$W^\a$ being the components of the ordinary spatial velocity of the PR frame with respect to the Zermelo frame. They satisfy
\begin{equation}
\label{spacvel}
 W_\alpha = \delta_{\alpha \beta}W^\beta, \quad  W^\alpha W_\alpha=W^i W_i=1-\frac{1}{\gamma^2}\, .
\end{equation}
In other words, at each spacetime point where an inertial observer meets a non-inertial one, $W^\alpha$ are the components of their relative velocity and $\gamma$ their relative Lorentz factor.  
For KAdS we find 
\beq\label{WK}
W^\a\check{z}_\a
=-\frac{a}{L}\frac{\sin\theta}{\sqrt{\Xi}} \check{z}_2
= -\frac{a}{L^2}\pa_\phi\,\quad \gamma=\sqrt{\frac{\Delta_\theta}{\Xi}}\,.
\eeq
The coordinate components are constant and hence the stationary
metric (\ref{KAdS-zer}) on the boundary can be made conformal to the static metric (\ref{KAdS-einrot}) by a linear diffeomorphism, as was noticed in \cite{Caldarelli:1999xj}. This feature is not present for TNAdS, in which case we have
\beq\label{WTN}
W=\frac{1}{\sqrt{\frac{L^2}{4n^2}\cot^2 \nicefrac{\theta}{2}-1}} \check{z}_2=\frac{n}{L^2 \cos^2 \nicefrac{\theta}{2}-4n^2  \sin^2 \nicefrac{\theta}{2}}\pa_\phi\, \quad \gamma=\frac{1}{\sqrt{1-\frac{4n^2}{L^2}\tan^2\nicefrac{\theta}{2}}}\,.
\eeq

Using Eqs. \eqref{coframeRanders}, \eqref{frameRanders}, \eqref{lorfac} and \eqref{h_lambda}, one can provide a complete and explicit  form of the Lorentz transformation \eqref{P-RZtransf} relating the PR and Zermelo  frames, whose temporal component is given in  \eqref{uzerm2}. One finds
\beq
\label{LambdaZerm}
\hat{z}= \Lambda \hat{e}\, , \quad \Lambda=\left(\begin{matrix}
      \gamma  & -\Gamma_\alpha^{\hphantom{\alpha}\gamma}W_\gamma  \\ -\gamma  W^\beta & \Gamma_\alpha^{\hphantom{\alpha}\gamma}\!\left[W_\gamma W^\beta+\frac{\delta_\gamma^\beta}{\gamma^2}\right]
    \end{matrix}
  \right),
\eeq
where 
\begin{equation}
\label{Gam}
 \Gamma_\alpha^{\hphantom{\alpha}\beta} =\gamma ^2 E_\alpha^{\hphantom{\alpha}i}L^\beta_{\hphantom{\beta}i}\, .
\end{equation}
Using the above relations it is straightforward to check that
\begin{equation}
\label{lorentz} 
\Lambda^{\mathrm{T}}\eta \Lambda=\eta\,.
\end{equation}

The Zermelo frame  (\ref{frame_Zermelo}) is non-inertial: there is an acceleration $\nabla_{\check{z}_0}\check{z}_0\neq 0$. In this frame, the fluid is non-static and its velocity changes,
\beq
\label{Zermacc}
\nabla_{\check{z}_0}\check{u}
=\omega_{0\alpha}^{\mathrm{Z}}\delta^{\alpha\beta}\check{z}_ \beta\, ,
\eeq
according to the vorticity as it is observed in the Zermelo frame: $\omega_{ab}^{\mathrm{Z}}=L_\alpha^{\hphantom{\alpha}i}L_\beta^{\hphantom{\beta}j}\omega_{ij}$ and $ \omega_{0\beta}^{\mathrm{Z}}=W^\alpha\omega_{\alpha\beta}^{\mathrm{Z}}$. The fluid's velocity vector $\check{u}$ undergoes a precession around the worldline of a Zermelo observer. The latter being an accelerated observer, the variation of $\check{u}$ is actually better captured as a Fermi derivative along $\check{z}_0$:
\beq
\label{Fermi_accel_Zerm}
\mathrm{D}_{\check{z}_0}\check u=\left( \omega_{0\alpha}^{\mathrm{Z}}-\check{z}_\alpha(\gamma)\right)\delta^{\alpha\beta}\check{z}_\beta + W^\alpha \check{z}_\alpha(\gamma)\check{z}_0\, ,
\eeq
where $\check{z}_\alpha(\gamma)= L_\alpha^{\hphantom{\alpha}i}\partial_i\gamma$. The extra terms result from the acceleration  of the Zermelo frame and contribute to the observed precession of the velocity vector $\check{u}$.  Contrary to the vorticity $ \omega_{0\alpha}^{\mathrm{Z}}$, which is generally non-zero in the PR backgrounds, the combination $ \omega_{0\alpha}^{\mathrm{Z}}-\check{z}_\alpha(\gamma)$ vanishes under the necessary and sufficient condition 
\beq
\label{ZAMO}
W^j \omega_{ji} = \gamma \partial_i \gamma\, ,
\eeq
also implying $W^\alpha \check{z}_\alpha(\gamma)=0$. Hence, under \eqref{ZAMO} the Fermi derivative \eqref{Fermi_accel_Zerm}  vanishes. In that case the effective precession of the fluid worldline with respect to the Zerrmelo observer disappears as a consequence of the cancellation of the genuine vorticity and of the effect produced by the acceleration. Equation  \eqref{ZAMO} carries an intrinsic information on the background: when fulfilled, the Zermelo observers coincide with the locally non-rotating (or ZAMO) frames  \cite{Bardeen}. Remarkably, this occurs for KAdS but not for TNAdS. 

\subsection{Closed timelike curves and optical horizons}

The emergence of closed timelike curves was mentioned  at the end of Sec. \ref{KAdS}, in the framework of the Som--Raychaudhuri metric (\ref{KAdS-bry-pol}). As CTCs appear also in the TNAdS geometry, they deserve a comprehensive discussion. 

Geometries of the general PR form may not be globally hyperbolic. In the Randers frame, where the metric assumes the form (\ref{Papa}), this happens whenever regions exist where $b^2=b_i b_j a^{ij}>1$. Indeed, in these regions, the spatial metric $ a_{ij} - b_i b_j$ possesses a negative eigenvalue, and constant-$t$ surfaces are no longer spacelike.  Therefore the extension of the physical domain accessible to the inertial observers moving along $\check u= \partial_t$ is limited to spacelike disks in which  $b^2<1$ holds. Within these regions, the classical fluid dynamics is consistent. 

The lack of hyperbolicity is similarly revealed as the breakdown of the Randers versus Zermelo relationship. Following (\ref{lorfac}) and (\ref{h_lambda}), we observe that $b^2$ is the norm of the fluid spatial velocity with respect to the Zermelo frame. On the edge of the spacelike physical surface\footnote{This edge is called \emph{velocity-of-light surface} in \cite{Gibbons}, where many interesting features and illustrative examples are exhibited.}, where $b^2=1$, $\lambda$ vanishes \emph{i.e.} the Lorentz factor relating  the comoving (Randers) and the Zermelo frames diverges: the fluid  and all its comoving observers reach the speed of light with respect to the latter frame.

The two cases under consideration in the present work are fundamentally different from the above viewpoint. On the one hand, in KAdS geometry $b^2=\frac{a^2\sin^2\theta}{L^2-a^2\cos^2\theta}$, which is bounded by 1 as long as $a<L$. On the other hand, for TNAdS $b^2=1$ when $\theta$ reaches $\theta_*=2\arctan \nicefrac{L}{2n}$. Hyperbolicity holds in the disk $0<\theta<\theta_*$, whereas it breaks down in the complementary disk ($\pi>\theta>\theta_*$) centered at the Misner string.

The breaking of hyperbolicity is usually accompanied with the appearance of CTCs. These are ordinary spacelike circles, lying in constant-$t$ surfaces, which become timelike when these surfaces cease being spacelike, \emph{i.e.}  when $b^2>1$. These CTCs differ in nature from those due to compact time (as in the $SL(2,\mathbb{R})$  group manifold), and cannot be removed by unwrapping time. They require an excision procedure for consistently removing the $b^2>1$ domain, in order to keep a causally safe spacetime, similar to what happens in the case of the three-dimensional Ba\~nados--Teitelboim--Zanelli black hole -- although in the latter case the trouble is not due to hyperbolicity issues.

Several further comments are in order for the above presentation to be complete when considering the case of TNAdS. There, the circles tangent to $\partial_\phi$ become CTCs for $\pi>\theta>\theta_*$. As explained in detail in Sec. \ref{TNADS}, the TNAdS boundary geometry (\ref{anagoed}) is a squashed three-sphere, 
homogeneous and axisymmetric\footnote{As $t$ is non-compact, these statements should be considered with care, because of the presence of the Misner string, which removes a point from the boundary geometry.  Bearing this point, everything is consistent.}. Homogeneity implies that CTCs are present everywhere, passing through any arbitrary point of spacetime. In the disk $\pi>\theta>\theta_*$, these are circles at constant $t$ centered around the Misner string; for  $0<\theta<\theta_*$, the CTCs are sections of cylinders normal to the constant-$t$ surfaces. The time coordinate $t$ evolves periodically along these elliptically shaped CTCs. 

The situation described here for the boundary of TNAdS, is generic for all three-dimensional homogeneous spacetimes. The case of Som--Raychaudhuri (Bianchi II) mentioned earlier and the celebrated G\"odel space (Bianchi VIII) are illustrative examples of how homogeneity combined with rotation necessarily leads to the breakdown of hyperbolicity and the emergence of CTCs.
G\"odel space in particular was the first to be recognized as plagued by CTCs. The CTCs present in these spaces, however, are not geodesics \cite{rebPRD83, Drukker:2003mg}. Their presence is therefore harmless for classical causality. This is why G\"odel-like solutions like the TNAdS boundary have never been truly discarded, leaving open the possibility of quantum mechanical validity\footnote{Attempts, among others in string theory within holography, were proposed a few years ago (see e.g.  \cite{Rooman:1998xf, Hikida:2003yd, Drukker:2003mg,Israel:2003cx,Israel:2004vv, Israel:2004cd} and references therein).}. 

Even though the boundary spacetime of TNAdS is homogeneous, the constant-$t$ surfaces are not.  Inertial observers, comoving with the fluid have therefore a different perception  depending on whether they are at $0<\theta<\theta_*$ or in the disk $\pi>\theta>\theta_*$, surrounding the Misner string. This gives a physical existence to the $b^2=1$ edge\footnote{Homogeneity implies, however, that another time, say $t'$ -- and thus another frame \{$\partial_{t'}, \partial_{\phi'}, \partial_{\theta'}$\} -- can be chosen such that constant-$t'$ surfaces are spacelike on another disk partly covering  $\pi>\theta>\theta_*$. This amounts to simultaneously moving the Misner string, while allowing inertial observers passing through $\pi>\theta>\theta_*$ to define their spacelike physical surface at constant $t'$.}, the meaning of which is better expressed in the Zermelo frame. In the latter, the fluid becomes superluminal and the Misner string is interpreted  as the core of the vortex with homogeneous vorticity.

The various troublesome features which appear in G\"odel-like spaces as the one at hand for the TNAdS boundary,
are intimately related with non-trivial rotational properties combined with homogeneous character. In other words, for the TNAdS boundary, they are due to the existence of a monopole-like Misner vortex\footnote{Since the bulk theory is such that the boundary does not have access to a charge current, the Misner vortex cannot be associated with a vortex in an ordinary superfluid, but is related to the spinning string of \cite{Mazur:1986gb}, the metric of which, Eq. (\ref{AKT-south}), indeed appears when zooming in on the southern pole.}. 
Although no satisfactory physical meaning has ever  been given to G\"odel-like spaces, the causal consistency of the latter being still questionable, they seem from our holographic perspective to admit a sensible interpretation in terms on conformal fluids evolving in homogeneous vortices.

The above discussion holds in the perspective of interpreting the holographic data as a genuine stationary fluid. There is however an alternative viewpoint already advertised, consisting in the analogue gravity interpretation of the boundary gravitational  background. From the latter, the physical data are still ($h_{ij}, W^i$) \emph{i.e.} a two-dimensional geometry and a velocity field.
However, their combination into (\ref{Zermelo}) is not a physical spacetime. The would-be light cone, in particular, is narrowed down to the sound or light velocities in the medium under consideration -- necessarily smaller than the velocity of light in vacuum. Consequently, the breaking of hyperbolicity or the appearance of CTCs are not issues of concern, and the regions where $\gamma$ becomes imaginary keep having a satisfactory physical interpretation as portions of space, where the medium is supersonic/superluminal  with respect to the sound/light velocity \emph{in the medium and not in the vacuum}. Finally, the virtual spacetime (\ref{Zermelo}) governs the mode propagation through the fluid. This way of thinking opens up a new chapter that requires adjusting suitably the standard holographic dictionary. The latter provides indirect information on the physical system that must be retrieved. This is under investigation.

\section{Inertial frames and rotational Hall viscosity}\label{visc}

We have devoted considerable effort in discussing the role of the observer's frame in our approach to holographic hydrodynamics. Here we will show that this understanding leads to the determination of rotational Hall viscosity for neutral rotating fluids in three dimensions. This non-dissipative transport coefficient is known to occur in a variety of physical systems with broken T-invariance. For example, in elastic media it occurs in topologically non-trivial states in the presence of a magnetic field or a fermionic gap \cite{Avron:1995fg,HLF}, and in finite-temperature hydrodynamics, it arises in the presence of magnetic fields \cite {Saremi:2011ab}. Here it arises in neutral fluids, with the T-breaking supplied by the vorticity of the fluid. The Hall viscosity is the ``gravitational'' analogue  of the Hall conductivity, in that it may be extracted from correlators of the stress--energy tensor. We show below that in this context there is a {\it classical} contribution to the Hall viscosity, analogous to the classical Hall conductivity that follows from Lorentz invariance in a medium with non-zero charge density in a uniform magnetic field. As far as we have been able to understand, this concept is not included in the recents works of parity-broken hydrodynamics in three dimensions \cite{Kovtun,Minwalla}.

As we have reviewed above, relativistic fluids are built using two generally independent ingredients, a coframe $\{\hat e^a\}$ describing the underlying geometry and a velocity field $\check{u}$, which at least locally and under suitable circumstances, defines a spacelike foliation. Holographic hydrodynamics also of course  uses these two ingredients to construct the boundary fluids. We parameterize the foliation locally via $\check{u}$,
along with the dual 1-form $\hat u$,
which is normalized as $\hat u(\check{u})=-1$. Then, the stress current one-form of a three-dimensional relativistic system\footnote{We use a different symbol for the stress current in this section so as to not confuse it with the corresponding holographic quantity defined in (\ref{fT}).} is given by 
\beq
\label{Jcur}
\hat J^a=\varepsilon u^a \hat u+ p(\hat {e}^a+u^a\hat{u}) -u^a\hat q-q^a\hat a+\hat t^a\,,
\eeq
with $\varepsilon$, $p$ the energy density and pressure. The first and second terms are respectively longitudinal and  transverse, and correspond both to the perfect fluid. Terms involving 
$\hat q$ are mixed because   $\hat q(\check{u})=0$, whereas the last term is purely transverse: 
$\hat t^a(\check{u})=0$. The terms with  $\hat q$ and $\hat t^a$ are viscous: 
$\hat q$ is the heat current and $\hat t^a$ is a transverse stress current. One can always chose a frame (\emph{i.e.} define the velocity field of the fluid) in such a way that the heat current vanishes. This frame is called the Landau frame and in that case the transverse stress current $\hat t^a$ encodes all dissipative and non-dissipative transport coefficients,  given as a derivative expansion of the velocity field $\check{u}$.  At lowest order in this expansion and in the Landau frame,
 \beq \label{Lanfr}
\hat J^a=\kappa \hat f^a+\kappa \hat f^a_{\mathrm{visc.}}\, ,
\eeq
with $f^a$ and $ f^a_{\mathrm{visc.}}$ respectively given in \eqref{bmomentum} and \eqref{Tvisc}.

Before treating the neutral fluids and the corresponding Hall viscosity, let us first review the simpler case of the classical Hall conductivity of a charged fluid in a homogenous magnetic field. We consider a system having a charge current one-form given by 
\beq
\hat J=\rho \hat u+\hat j\,,
\eeq
where $\rho$ is the charge density and $\hat j$ is the transverse dissipative part,  $\hat j(\check{u})=0$.
The system is in a homogenous magnetic field which we define with respect to a spatial frame transverse to $\hat u$, $F=\frac12 B\epsilon_{\alpha\beta}\hat{e}^\alpha\wedge \hat{e}^\beta$, where $B$ is constant. To extract the conductivity, we turn on a small electric field
\beq
F=\frac12 B\epsilon_{\alpha\beta}\hat{e}^\alpha\wedge \hat{e}^\beta - {\cal E}_\alpha \hat{u}\wedge \hat{e}^\alpha\,
\eeq
and consider the transverse current induced,
\beq\label{eq:defHallcond}
 J_\alpha=\sigma_{\mathrm{H}} \varepsilon_{\alpha\beta}{\cal E}^\beta\,.
\eeq 
The Hall conductivity, $\sigma_{\mathrm{H}}$ can be extracted by noting that there is a small local Lorentz transformation to a frame in which the electric field vanishes. For example, if we take ${\cal E}_\alpha = ({\cal E}_1,0)$, then one can check that the relevant boost is in the 2-direction, with velocity $v=\nicefrac{{\cal E}_1}{B}$. From the point of view of the original frame, the boosted frame moves with velocity $v$, and the charge density then gives rise to a current 
\beq
J^2=\rho v=\frac{\rho}{B} {\cal E}_1\,.
\eeq
By comparison to (\ref{eq:defHallcond}), we derive $\sigma_{\mathrm{H}}=\nicefrac{\rho}{B}$. 

The same result arises from an argument that appeals to the  Lorentz force. The idea here is that if an electric field is turned on we can still get equilibrium (zero net force) if there is a non dissipative current induced, but the dissipative part of the current vanishes. To do so, we have to use extra dynamical information regarding the force on a current in an electromagnetic field. The latter written as a one-form equation is
\beq
m\hat a=
F(\cdot,\check{J})\,,
\eeq
where $F$ is the electromagnetic field strength and 
where we define the acceleration as above
$
\hat{a}=\nabla_{\check{u}}\hat{u}
$.
So we find the components of the force
\beq
 F(\check{e}_\alpha,\check{J})=\frac12 B\epsilon_{\alpha\beta}J^\beta+\frac12\left( {\cal E}_\alpha \hat u(\check{J})-{\cal E}_\beta \hat u(\check{e}_\alpha)J^\beta\right)=\frac12 \left(\rho-B\sigma_{\mathrm{H}}\right){\cal E}_\alpha\, .
 \eeq
 The vanishing of the force implies that $\sigma_{\mathrm{H}}=\nicefrac{\rho}{B}$. This effect can also be thought of in terms of a current--current correlator -- the conductivity being computed in linear response by finding the current induced by turning on a small electric field:
\beq
\sigma_{\mathrm{H}}\sim\lim_{\omega\to 0}\frac{1}{\omega}\langle J_x J_y\rangle_{\omega}\, .
\eeq
One way to think of this induced current is that if we insist on writing $\hat J=\rho \hat u$, it is as if we have modified $\check{u}$ to have a piece along the foliation. 

Our result for neutral fluids follows a direct generalization of the above logic. Consider a non-accelerating neutral fluid with uniform vorticity 
\beq\label{univort}
\omega=\Omega\hat{e}^1 \wedge \hat{e}^2,
\eeq
where $\Omega$ is constant. 
The analogue of turning on an electric field would be to modify the frame (or equivalently deform the  metric). This will clearly lead by linear response to the stress--stress correlator. 
Define
\beq
 \langle J^a_\mu J^b_\nu\rangle_q= \cdots+i\zeta_{\mathrm{H}} \eta^{ab}\epsilon_{\mu\nu\lambda}q^\lambda\,,
 \eeq
 where $\{q^\mu\} = \{\omega,q^1,q^2\}$ is the 3-momentum.
Let us focus in particular on the  correlator
\beq
\langle J^0_x J^0_y\rangle_q= \cdots-i\zeta_{\mathrm{H}} \det \hat e\ \omega\,.
\eeq
In linear response, this means that
\beq
\left.\delta J^a_\rho\right|_{\delta  \hat e} = \cdots+i\zeta_{\mathrm{H}} g_{\mu\rho}\epsilon^{\mu\nu\lambda}q_\lambda \delta e^a_\nu 
\eeq
or as a form\footnote{In elastic solids, $\delta \hat J^a=\zeta_{\mathrm{H}} *\delta T^a$, with $T^a$ the torsion 2-form \cite{HLF}. For simplicity, here we do not modify the spin connection.}, $\left.\delta \hat J^a\right|_{\delta  \hat e}=\zeta_{\mathrm{H}} *\mathrm{d}\delta  \hat e^a\equiv \zeta_{\mathrm{H}}\delta \hat{\cal E}^a$, where $*\hat{\cal E}^a$ is the analogue of the electric field in the magnetic problem discussed above. Here we will focus on $\hat J^0$, and thus turn on a small $\hat{\cal E}^0$. Thus, for example, if we turn on ${\cal E}^0_x$, we expect to see a contribution to $J^0_y$. Consider a small variation in the frame of the form $\hat e^0=-\hat u+\delta \hat e^0$, with $\delta\hat e^0=\delta e^0_x(t)\mathrm{d}x$. We then obtain
\beq
*\hat{\cal E}^0=\Omega \mathrm{d}x\wedge \mathrm{d}y-(\pa_t\delta e^0_x)\mathrm{d}x\wedge \mathrm{d}t.
\eeq
We note that a small linear diffeomorphism with velocity
\beq
v_y=\frac{\pa_t\delta e^0_x}{{\Omega}}
\eeq
will bring us back to the rest frame, in which $*\hat{\cal E}^0=\Omega \mathrm{d}x\wedge \mathrm{d}y$.
In the original frame of reference, we then see a stress current $\delta J^0_y=(\varepsilon+p)v_y$ (obtained by inserting $\hat u=\hat e^0+\hat v$ into the perfect fluid stress current). We conclude that
\beq
\label{zetaH}
\zeta_{\mathrm{H}}=\frac{\varepsilon+p}{\Omega}
\,.
\eeq
We can also understand this effect as a force balancing, and we explore a variety of alternative derivations of our result in Appendix \ref{appA}. 

The holographic derivation of our result (\ref{zetaH}) would require to use
the TNAdS geometry discussed in the previous sections, since the corresponding boundary geometry exhibits precisely a uniform vorticity. The determination of the Hall viscosity is then expected to arise from the calculation of the $\langle TT\rangle$ correlator in a manner similar to the emergence of the classical Hall conductivity using a dyonic $\mathrm{AdS}_4$ black hole \cite{Kovtun2}.  

In the KAdS case the vorticity is non-uniform and actually vanishes at the equator. One might think that the result in this case would be also (\ref{zetaH}) with $\Omega$ being the $\theta$-dependent local vorticity. This would require however a more involved computation to be set (e.g. see \cite{Sonner}). 

It is remarkable that our formula (\ref{zetaH}) is similar to the classical Hall viscosity coefficient of magnetized plasmas \cite{Landau}
\beq
\label{zetaHL}
\zeta_{\mathrm{H}}^{L}=\frac{Nk_BT}{2\omega_c}\,,\quad \omega_c=\frac{qB}{m}\,,
\eeq
where $N$ is the density of particles with charge $q$ and mass $m$ that constitute the plasma. The relationship of (\ref{zetaHL}) with the Hall viscosity of quantum systems was explained in \cite{Read}.  

The quantum version of our result (\ref{zetaH}) is not yet known. However, it is natural to expect that it will emerge\footnote{A.C.P. thanks N. Cooper for a discussion on this point.} following the systematic analogy between rotating bosons and quantum Hall systems \cite{Nigel}. We will not consider this interesting problem further in this work.

\section{Summary and outlook}\label{sum}

In this work we have initiated the holographic description of rotating fluids having in mind possible applications to rotating bose gases and also analogue gravity systems. Following a 3+1-split holographic formalism we were able to clarify the important issue of the frame from which the boundary fluid is observed. We find that the Fefferman--Graham expansion corresponds to the fluid's description from a comoving inertial Papapetrou--Randers frame. However, the physical picture of the boundary fluid as a moving medium becomes manifest in the non-inertial Zermelo frame, which can be reached by a suitable Lorentz transformation. 

In the  Zermelo picture, the metric has the form of the acoustic/optical geometries that arise in studies of light and sound propagation in moving media. This picture provides therefore an alternative interpretation of our set up: either the background metric is the physical spacetime metric experienced by the fluid; or the analogue metric associated with the propagation of fluctuations in some other medium. In the latter case, it is possible to  identify regions in the fluid that move with superluminal/supersonic velocities.

We apply our general results to the cases of the Kerr--AdS$_4$ and Taub--NUT--AdS$_4$ geometries and show that they describe fluids in cyclonic and vortex motions respectively. In the TNAdS case the homogeneity of the vorticity leads to velocity-of-light surfaces which we interpret as boundaries for superluminal rotation of the fluid, with respect to a specific observer. This is physically acceptable in the analogue picture, since in that case the velocity of light in the boundary is smaller than the vacuum velocity of light. 

Finally, we use our understanding of the observer's frame in hydrodynamics to calculate the classical Hall viscosity for three-dimensional rotating fluids. We were not able to find this result in recent works on parity non invariant hydrodynamics although it might be there.

Our work opens up various avenues for exploration. The first and obvious think to do is to study fluctuations around our geometries and in particular around TNAdS in order to calculate holographically the transport properties of our boundary fluids. This project is in its final stages and the results will be presented soon \cite{LPP_new}. It would be also interesting to study the relationship of our approach to other approaches of holographic hydrodynamics such as \cite{Kovtun,Minwalla} and references therein. In this line, we could ask the reverse question namely whether one can use our approach to study holographic fluids with preordered vorticity. As we have seen, the Kerr and TN fluids are cyclones and vortices, but one could ask how could more complicated flows could be described holographically. This can be formulated as the question of studying the generalization of Weyl's multipole solutions to asymptotically AdS$_4$ spaces. We hope to report on this issues in the forthcoming work \cite{CLPPPS}. Finally, our spinoff result for the rotational Hall viscosity and its quantum counterpart deserve further study and comparison with theoretical and experimental work on rotating Bose gases.

\section*{Acknowledgements}
The authors benefited from discussions with  C. Bachas, M. Caldarelli, C. Charmousis, T. Hughes, R. Meyer, V. Niarchos, G. Policastro, V. Pozzoli, K. Sfetsos, K. Siampos. P.M.P. would like to thank the University of Crete and the University of Patras, and R.G.L. and A.C.P. thank the CPHT of \'Ecole Polytechnique for hospitality. 
This research was supported by the LABEX P2IO, the
French ANR contract  05-BLAN-NT09-573739 \textsl{String cosmology}, the ERC Advanced Grant  226371 \textsl{Mass hierarchy and particle physics at the TeV scale},
the ITN programme PITN-GA-2009-237920 
\textsl{Unification in the LHC era}, the IFCPAR CEFIPRA programme 4104-2 \textsl{Moduli stabilization, magnetized branes and particles}, and the U.S. Department of Energy contract FG02-91-ER4070. The work of A.C.P.  was in part supported by the European grant
FP7-REGPOT-2008-1 \textsl{CreteHEPCosmo-228644}. 
 
\begin{appendix}

\section{Classical rotational Hall viscosity in metric language}\label{appA}

In Sec. \ref{visc}, we presented the determination of the Hall viscosity using form language in arbitrary frame. For the reader's convenience, we supplement here the computation in metric language.  We will first present the paradigm of the Hall conductivity for a charged fluid and then proceed with the analysis of classical Hall viscosity in a neutral fluid of the type we have been analyzing in the main part of the paper. 

In order  derive the classical Hall conductivity in metric language consider a charged fluid in stationary motion,  whose current density  reads:
\beq
J^\m =\rho u^\m\, ,
\eeq
where $\rho$ is the constant charge density. Suppose that the fluid couples to an external source with  field strength $F_{\m\n}=\partial_\m A_\n - \partial_\n A_\m$,  such that its equilibrium is unaffected. The necessary condition for that is to require the exerted Lorentz force on the current be zero
\beq
F^{\m}_{\hphantom{\m}\n}J^\n=0\, .
\eeq
Suppose now that the external source is perturbed as
\beq
F^{\m}_{\hphantom{\m}\n}\to F^{\m}_{\hphantom{\m}\n} +\delta F^{\m}_{\hphantom{\m}\n}\, .
\eeq
As a response, the velocity field of the fluid must be modified accordingly in order to respect the zero force condition and keep the flow in the non-dissipating, non-forced stationary state. This response of the fluid satisfies therefore
\beq\label{zerofLor}
\delta F^{\m}_{\hphantom{\m}\n}u^\n +F^{\m}_{\hphantom{\m}\nu}\delta u^\n=0\, .
\eeq
In the comoving frame $\{u^{\m}\}=\{1,0,0\}$ and last condition is recast as
\beq
F^{\m}_{\hphantom{\m}\n}\delta J^\n = -\rho \delta F^{\mu}_{\hphantom{\mu} t}\, 
\eeq
where $\delta J^\n=\rho \delta u^\n$ is the non-dissipating, spatial current-density deformation response seen by the no longer comoving observer.
Considering a constant magnetic field of the form  $F^{i}_{\hphantom{i}j}=\epsilon^{i}_{\hphantom{i} j}B$ as the initial background and 
electric perturbations of the kind 
\beq \label{pert-curv-F}
\delta F^{i}_{\hphantom{\mu} t}=i\omega\delta A^i
\eeq 
(in the gauge $A_0=0$), one produces a spatial response: 
\beq
\delta J^i = -i\omega \frac{\rho}{B}\epsilon^{i}_{\hphantom{i} j}\delta A^j\, .
\eeq
The proportionality coefficient between the external electric perturbation and the induced current is the Hall conductivity. This can also be obtained from the usual Kubo formula
\beq
\sigma_{\mathrm{H}}=\frac{\rho}{B}=-\lim_{\omega\rightarrow 0}\frac{1}{2\omega}{\rm Im}\left\langle J^i\left(\omega,\vec{0}\right)J^j\left(\omega,\vec{0}\right)\right\rangle\epsilon_{ij}\, .
\eeq
En passant, the above short calculation shows that when $B=0$, non-dissipative spatial current  cannot emerge as compensator for possible fluctuations of the external electric field. The singularity appearing in that case is therefore spurious.
 
The above reasoning can be applied to the neutral fluid, which is the main purpose of our paper.  In this case, the role of the background and external source is played by the metric, whereas the velocity field encodes the response. Indeed, relativistic fluids are described  using the background metric $g_{\m\n}$ and the velocity field (normalized as $u^\m u_\m =-1$). 
From these two quantities one constructs the longitudinal and normal projectors introduced in \eqref{proj}
and decompose the energy--momentum tensor as described in Secs. \ref{emhol} and \ref{visc}, Eqs. \eqref{bmomentum}, \eqref{Tvisc} and \eqref{Lanfr}. The results are summarized as follows:
\beq \label{emlan}
T^{\m\n}=\varepsilon u^\m u^\n +p  h_{\m\n}+t_{\m\n}\, ,
\eeq
where $\varepsilon$ and $p$ are the energy density and pressure.
At lowest order in the velocity derivative expansion one finds 
\begin{equation}\label{Sigvisc}
t_{\m\n}=-2\eta \sigma_{\m\n}-\zeta h_{\m\n}\Theta+\zeta_{\mathrm{H}} \epsilon^{\vphantom{\rho}}_{(\m\lambda\rho} u^\lambda\sigma^\rho_{\hphantom{\rho}\n)}
\end{equation}
with $\eta, \zeta, \zeta_{\mathrm{H}} $ the shear, bulk and Hall viscosities.
Note that expressions  \eqref{emlan} and \eqref{Sigvisc} assume in general being in the Landau frame, where the spatial momentum density (also called the heat current) vanishes. If one slightly deviates from this assumption, one can recover it by an appropriate velocity shift
\beq
u^\m\rightarrow u^\m +\delta u^\m \,,\quad \delta u^\m = \frac{1}{\varepsilon +p}\delta q^\m\, .
\eeq
In the present context, will not consider situations with non-vanishing heat current.

The hydrodynamic equations follow from $\nabla_\m T^{\m\n}=0$. In thermal equilibrium, the energy density and pressure  are related to the entropy density $s$ and local temperature as $\varepsilon +p=sT$ (the fluid is neutral and its chemical potential vanishes). As mentioned earlier, for a conformal fluid the energy--momentum tensor has vanishing trace, which implies the equation of state $2p=\varepsilon$ and zero bulk viscosity. 

When turning on a gravitational field as a background of a stationary flow -- starting assumption here as in the previously studied charged fluid -- the zero force condition \eqref{zerofLor} translates into 
\beq\label{eq:geodesiczero}
\Gamma^{\m}_{\n\rho}u^\n u^\rho = 0\, .
\eeq
Imposing that upon a metric disturbance acting as a source, the zero-force condition remains unaltered, leads to the following relationship with the induced perturbation on the velocity field -- the response:
\beq
\label{B12}
u^\n u^\rho\delta\Gamma^\m_{\n\rho} +2\Gamma^\m_{\n\rho}\delta u^\n u^\rho=0\,.
\eeq

Assuming a natural comoving frame ($\check{u}=\partial_t$),  condition \eqref{eq:geodesiczero} becomes $\Gamma^\m_{tt}=0$, and is automatically satisfied in Papapetrou--Randers geometries \eqref{Papa} (see footnote \ref{conne}), which we will assume being the gravitational background subject to perturbation. Under these conditions, \eqref{B12} reads:
\beq
\label{B13}
\delta\Gamma^\m_{tt} =-2\Gamma^\m_{tk}\delta u^k\,.
\eeq
or, using the actual expressions for the connection coefficients in terms of the vorticity components $\omega_{ij}$, 
\beq
\label{B13PA}
\delta\Gamma^\m_{tt} =2g^{\m j} \omega_{jk}\delta u^k\, ,
\eeq

We have so far expressed the perturbation induced on the velocity field, $\delta  u^k$, as resulting from a perturbation on the connection, $ \delta\Gamma^\m_{tt}$. One can express the latter results directly in terms of the metric perturbation,  $\delta g_{\m\n}$, using the definition of the Levi--Civita coefficients. With the gauge choice\footnote{The gauge choice $g_{tt}=-1$ together with $\check{u}=\partial_t$ implies that $u^t=1, u^i=0, u_t=-1$ and  $u_i=g_{it}$, generally non-vanishing. Since it is assumed to also hold after the perturbation ($\delta g_{tt}=0$), it also implies that $\delta u^i g_{it=0}$. The gauge fixing can be supplemented with $\delta u^t=0$, which further leads to $\delta u_t=0, \delta u_j= \delta g_{jt} +g_{jk}\delta u^k$ whereas $\delta u^k$ is left free \emph{i.e.}  to be determined dynamically.} $g_{tt}=-1$, we obtain 
\beq\label{delgam-u}
\delta \Gamma^\m_{tt} = i\omega g^{\m \n}\delta g_{\n t}\,,
\eeq
which generalizes \eqref{pert-curv-F} in the gravitational case. Combined with \eqref{B13PA} it leads to\footnote{Notice that, $ \delta \Gamma^t_{tt}=
g_{t j} \delta \Gamma^j_{tt}\Leftrightarrow g_{t\m}\delta \Gamma^\m_{tt}=0 $. If the latter were non-vanishing, it would imply  $\delta g_{tt}\neq 0$, in contradiction with  the gauge condition $ g_{tt}=-1$.}
\beq\label{delg-u}
2\omega_{jk}\delta u^k = i \omega \delta g_{jt}
\,.
\eeq
Setting finally 
\beq\label{omegcompo}
\omega=\Omega\mathrm{d}x^1 \wedge \mathrm{d}x^2,
\eeq
as in \eqref{univort}, we obtain:
\beq\label{delu}
\delta u^i = \frac{\omega}{2i\Omega}\epsilon^{ij} \delta g_{jt}
\,.
\eeq

In order to reach our goal, we would like to ultimately express the perturbation of the metric and the of velocity field in terms of the induced perturbation on the energy--momentum tensor. The latter is of the perfect-fluid type because we assumed the fluid to be originally in a free stationary motion, with uniform energy density and pressure, and the disturbances to preserve this property. Hence,
\beq
\label{delT}
\delta T^{\mu\nu}=
2(\varepsilon +p) \delta u^{(\m} u^{\n)} +p \delta g^{\mu\nu}\, ,
\eeq
leading in particular to (we use the gauge conditions)
\beq
\label{delT}
\delta T^i_{\hphantom{i}t}=-(\varepsilon +p) \delta u^i.
\eeq
Combining with \eqref{delu}, we find
\beq
\label{delT-fin}
\delta T^i_{\hphantom{i}t}=  \frac{i\omega}{2}  \frac{\varepsilon +p}{\Omega} \epsilon^{ij} \delta g_{jt},
\eeq
and we read from corresponding Kubo formula the rotational Hall viscosity coefficient as
\beq
\label{sigmaH}
\zeta_{\mathrm{H}}=\lim_{\omega\rightarrow 0}\frac{1}{\omega}{\rm Im}\left\langle T^{i}_{\hphantom{i}t}\left(\omega,\vec{0}\right)T^{j}_{\hphantom{j}t}\left(-\omega,\vec{0}\right)\right\rangle\epsilon_{ij}=\frac{\varepsilon +p}{\Omega}\,.
\eeq
One thus recovers \eqref{zetaH}.
\end{appendix}


\begin{thebibliography}{99}


\bibitem{Janik}
  R.A.~Janik,
 ``The dynamics of quark-gluon plasma and AdS/CFT,''
  Lect.\ Notes Phys.\  {\bf 828} (2011) 147
  [arXiv:1003.3291 [hep-th]].

\bibitem{HHH}
  S.A.~Hartnoll, C.P.~Herzog and G.T.~Horowitz,
  ``Building a holographic superconductor,''
  Phys.\ Rev.\ Lett.\  {\bf 101} (2008) 031601
  [arXiv:0803.3295 [hep-th]].


\bibitem{Esko}
  E.~Keski--Vakkuri and P.~Kraus,
``Quantum Hall effect in AdS/CFT,''
  JHEP {\bf 0809} (2008) 130
  [arXiv:0805.4643 [hep-th]].

\bibitem{Liu}
  H.~Liu, J.~McGreevy and D.~Vegh,
 ``Non-Fermi liquids from holography,''
  Phys.\ Rev.\ {\bf D83} (2011) 065029
  [arXiv:0903.2477 [hep-th]].
  
  

\bibitem{Hoyos}
  C.~Hoyos--Badajoz, K.~Jensen and A.~Karch,
  ``A holographic fractional topological insulator,''
  Phys.\ Rev.\ {\bf D82} (2010) 086001
  [arXiv:1007.3253 [hep-th]].
  
\bibitem{HLF}
  T.L.~Hughes, R.G.~Leigh and E.~Fradkin, ``Torsional response and dissipationless viscosity in topological insulators,''
  Phys.\ Rev.\ Lett.\  {\bf 107} (2011) 075502
  [arXiv:1101.3541 [cond-mat.mes-hall]].

\bibitem{Rangamani}
  V.E.~Hubeny, S.~Minwalla and M.~Rangamani,
  ``The fluid/gravity correspondence,''
  arXiv:1107.5780 [hep-th].
  
\bibitem{Kovtun}
  K.~Jensen, M.~Kaminski, P.~Kovtun, R.~Meyer, A.~Ritz and A.~Yarom,
 ``Parity-violating hydrodynamics in $2+1$ dimensions,''
  arXiv:1112.4498 [hep-th].
  
\bibitem{Minwalla}
  N.~Banerjee, J.~Bhattacharya, S.~Bhattacharyya, S.~Jain, S.~Minwalla and T.~Sharma,
  ``Constraints on fluid dynamics from equilibrium partition functions,''
  arXiv:1203.3544 [hep-th].

\bibitem{Son}
  D.T.~Son and P.~Surowka,
  ``Hydrodynamics with triangle anomalies,''
  Phys.\ Rev.\ Lett.\  {\bf 103} (2009) 191601
  [arXiv:0906.5044 [hep-th]].
  
\bibitem{Landsteiner}
  K.~Landsteiner, E.~Megias, L.~Melgar and F.~Pena--Benitez,
  ``Holographic gravitational anomaly and chiral vortical effect,''
  JHEP {\bf 1109} (2011) 121
  [arXiv:1107.0368 [hep-th]].


\bibitem{Cooper}
N.R. Cooper, ``Rapidly rotating atomic gases,''  Adv. in Phys. \textbf{57} (2008) 539.

\bibitem{Fetter}
A. L. Fetter, ``Rotating trapped Bose--Einstein condensates,'' 
Rev. Mod. Phys. \textbf{81} (2009) 647.

\bibitem{Dalibard}
M. Roncaglia, M. Rizzi and J. Dalibard
``From rotating atomic rings to quantum Hall states,''
Scientific Reports \textbf{1} (2011) 43.


\bibitem{Barcelo}
  C.~Barcel\`o, S.~Liberati and M.~Visser,
 ``Analogue gravity,''
  Living Rev.\ Rel.\  {\bf 8} (2005) 12
  [gr-qc/0505065].


\bibitem{Unruh:1980cg}
  W.G.~Unruh,
  ``Experimental black hole evaporation,''
  Phys.\ Rev.\ Lett.\  {\bf 46} (1981) 1351.

\bibitem{Unruh:1994je}
  W.G.~Unruh,
  ``Sonic analog of black holes and the effects of high frequencies on black
  hole evaporation,''
  Phys.\ Rev.\  {\bf D51} (1995) 2827.



\bibitem{Cacciatori}
 S.L.~Cacciatori, F.~Belgiorno, V.~Gorini, G.~Ortenzi, L.~Rizzi, V.G.~Sala and D.~Faccio,
 ``Space--time geometries and light trapping in traveling refractive index perturbations,''
  New J.\ Phys.\  {\bf 12} (2010) 095021
  [arXiv:1006.1097 [physics.optics]].

\bibitem{Liberati}
  S.~Liberati, A.~Prain and M.~Visser,
 ``Quantum vacuum radiation in optical glass,''
  arXiv:1111.0214 [gr-qc].

\bibitem{Shapere}
  S.R.~Das, A.~Ghosh, J.-H.~Oh and A.D.~Shapere,
 ``On dumb holes and their gravity duals,''
  JHEP {\bf 1104} (2011) 030
  [arXiv:1011.3822 [hep-th]].

\bibitem{Leigh:2007wf} 
  R.G.~Leigh and A.C.~Petkou,
  ``Gravitational duality transformations on (A)dS$_4$,''
  JHEP {\bf 0711}, 079 (2007)
  [arXiv:0704.0531 [hep-th]].

\bibitem{Mansi:2008br} 
  D.S.~Mansi, A.C.~Petkou and G.~Tagliabue,
  ``Gravity in the $3+1$-split formalism I: holography as an initial value problem,''
  Class.\ Quant.\ Grav.\  {\bf 26} (2009) 045008
  [arXiv:0808.1212 [hep-th]].
  
\bibitem{Mansi:2008bs} 
  D.S.~Mansi, A.C.~Petkou and G.~Tagliabue,
  ``Gravity in the $3+1$-split formalism II: self-duality and the emergence of the gravitational Chern--Simons in the boundary,''
  Class.\ Quant.\ Grav.\  {\bf 26}, 045009 (2009)
  [arXiv:0808.1213 [hep-th]].


\bibitem{Gibbons}
  G.W.~Gibbons, C.A.R.~Herdeiro, C.M.~Warnick and M.C.~Werner,
 ``Stationary metrics and optical Zermelo--Randers--Finsler geometry,''
  Phys.\ Rev.\ {\bf D79} (2009) 044022
  [arXiv:0811.2877 [gr-qc]].

\bibitem{Bardeen} 
J.M. Bardeen, ``A variational principle for rotating stars in general relativity,'' Ast. Journ. \textbf{162} (1970) 71. 

\bibitem{Leonhardt} 
U. Leonhardt and P. Piwnicki, ``Relativistic effects of light in moving media with extremely low group velocity,'' Phys.  Rev. Lett.  84 (2000) 822.

\bibitem{Read}
N. Read, ``Non-Abelian adiabatic statistics and Hall viscosity in quantum Hall states and $p_x+ip_y$ paired superfluids,''
Phys. Rev. \textbf{B79} (2009) 045308.


\bibitem{Caldarelli:2008mv}
  M.M.~Caldarelli, O.J.C.~Dias, R.~Emparan and D.~Klemm,
  ``Black holes as lumps of fluid,''
  JHEP {\bf 0904} (2009) 024
  [arXiv:0811.2381 [hep-th]].

\bibitem{Caldarelli:2008ze}
  M.M.~Caldarelli, O.J.C.~Dias and D.~Klemm,
  ``Dyonic AdS black holes from magnetohydrodynamics,''
  JHEP {\bf 0903} (2009) 025
  [arXiv:0812.0801 [hep-th]].

\bibitem{Papapetrou}
A. Papapetrou, ``Champs gravitationnels stationnaires \`a sym\'etrie axiale,''  Ann. Inst. H. Poincar\'e {\bf A4} (1966) 83.


\bibitem{Policastro}
  A.~Amariti, D.~Forcella, A.~Mariotti and G.~Policastro,
  ``Holographic optics and negative refractive index,''
  JHEP {\bf 1104} (2011) 036
  [arXiv:1006.5714 [hep-th]].
  
  
  \bibitem{Zer31}
E. Zermelo, ``\"Uber das Navigationsproblem bei ruhender oder ver\"anderlicher Windverteilung,'' Z. Angew. Math. Mech. \textbf{11} (1931) 114.


\bibitem{Shen}
Z. Shen, ``Finsler metrics with $K=0$ and $S=0$,'' Canadian J. Math. {\bf 55} (2003) 112 [arXiv:math/0109060].




\bibitem{Henneaux:1985tv}
  M.~Henneaux and C.~Teitelboim,
  ``Asymptotically anti-de Sitter spaces,''
  Commun.\ Math.\ Phys.\  {\bf 98} (1985) 391.

\bibitem{Hawking:1998kw}
  S.W.~Hawking, C.J.~Hunter and M.~Taylor,
  ``Rotation and the AdS/CFT correspondence,''
  Phys.\ Rev.\   {\bf D59} (1999) 064005
  [arXiv:hep-th/9811056].
  
\bibitem{Awad:1999xx}
  A.M.~Awad and C.V.~Johnson,
  ``Holographic stress tensors for Kerr--AdS black holes,''
  Phys.\ Rev.\   {\bf D61} (2000) 084025
  [arXiv:hep-th/9910040].

\bibitem{Caldarelli:1999xj}
  M.M.~Caldarelli, G.~Cognola and D.~Klemm,
  ``Thermodynamics of Kerr-Newman-AdS black holes and conformal field
  theories,''
  Class.\ Quant.\ Grav.\  {\bf 17} (2000) 399
  [arXiv:hep-th/9908022].

\bibitem{Gibbons:2004ai}
  G.W.~Gibbons, M.J.~Perry and C.N.~Pope,
  ``The first law of thermodynamics for Kerr-anti-de Sitter black holes,''
  Class.\ Quant.\ Grav.\  {\bf 22} (2005) 1503
  [arXiv:hep-th/0408217].

\bibitem{Gibbons:1979xm}
  G.W.~Gibbons and S.W.~Hawking,
  ``Classification of gravitational instanton symmetries,''
  Commun.\ Math.\ Phys.\  {\bf 66} (1979) 291.

\bibitem{Hunter:1998qe}
  C.J.~Hunter,
  ``The action of instantons with nut charge,''
  Phys.\ Rev.\   {\bf D59} (1999) 024009
  [arXiv:gr-qc/9807010].

\bibitem{Manko:2009xx}
  V.S.~Manko, E.D.~Rodchenko, E.~Ruiz and M.B.~Sadovnikova,
  ``Formation of a Kerr black hole from two stringy NUT objects,''
  arXiv:0901.3168 [gr-qc].
  
     \bibitem{SR68}
  M.M. Som and  A.K. Raychaudhuri, ``Cylindrically symmetric charged dust distribution in rigid rotation in general relativity,''Proc. R. Soc. London \textbf{A304} (1968) 81.
  
    \bibitem{rayPRD80}
  A.K. Raychaudhuri and S.N. Guha Thakurta,
  ``Homogeneous space--times of the G\"odel type,'' Phys. Rev. \textbf{D22} (1980) 802.
  
      \bibitem{rebPRD83}
M.J. Reboucas and J. Tiomno,
  ``Homogeneity of Riemannian space--times of G\"odel type,'' Phys. Rev. \textbf{D28} (1983) 1251.

      \bibitem{rebPLA87}
F.M. Paiva, 
M.J. Reboucas 
and A.F.F. Teixeira,
  ``Time travel in the homogeneous Som--Raychaudhuri universe,'' Phys. Lett. \textbf{A126} (1987) 168.

\bibitem{Taub:1950ez}
  A.H.~Taub,
  ``Empty spacetimes admitting a three parameter group of motions,''
  Annals Math.\  {\bf 53} (1951) 472.
  
\bibitem{NUT}
E.T. Newman, L. Tamburino and T.J. Unti,
``Empty-space generalization of the Schwarzschild metric,''
Journ. Math. Phys. \textbf{4} (1963) 915.

\bibitem{HE73} 
S.W. Hawking and G.F.R. Ellis,  \textsl{The large scale structure of space--time}, Cambridge University Press, Cambridge, 1973.
 
\bibitem{Eguchi:1978gw}
  T.~Eguchi and A.J.~Hanson,
  ``Selfdual solutions to Euclidean gravity'',
  Annals Phys. \textbf{120} (1979) 82.

\bibitem{Eguchi:1979yx}
  T.~Eguchi and A.J.~Hanson,
  ``Gravitational Instantons'',
  Gen.  Rel.  Grav.  \textbf{11} (1979) 315.

\bibitem{P85}
H. Pedersen, ``Eguchi-Hanson metrics with cosmological constant'', Class. Quantum Grav. \textbf{2}  (1985) 579.

\bibitem{Zoubos:2002cw}
  K.~Zoubos,
  ``Holography and quaternionic Taub--NUT,''
  JHEP {\bf 0212} (2002) 037
  [arXiv:hep-th/0209235].
  
 \bibitem{PVH12}
 P.M. Petropoulos and P. Vanhove, ``Gravity, strings, modular and quasimodular forms'', to appear in Annales Math\'ematiques Blaise Pascal. 
  
\bibitem{misner:1963}
C. Misner, ``The flatter regions of Newman, Unti and Tamburino's generalized Schwarzshild space,'' Jour. Math. Phys. \textbf{4} (1963) 924.

 \bibitem{bonnor:1969}
 W.B. Bonnor, ``A new interpretation of the NUT metric in general relativity,'' Proc. Camb. Phil. Soc. \textbf{66} (1975) 145.
 
 
\bibitem{dowker:1974}
J.S. Dowker, ``The NUT solution as a gravitational dyon,'' GRG \textbf{5} (1974) 603. 

\bibitem{Astefanesei:2004kn}
  D.~Astefanesei, R.B.~Mann and E.~Radu,
  ``Nut charged spacetimes and closed timelike curves on the boundary,''
  JHEP {\bf 0501} (2005) 049
  [arXiv:hep-th/0407110].

  
\bibitem{Rooman:1998xf}
  M.~Rooman and P.~Spindel,
  ``G\"odel metric as a squashed anti-de Sitter geometry,''
  Class.\ Quant.\ Grav.\  {\bf 15} (1998) 3241
  [gr-qc/9804027].

\bibitem{Hikida:2003yd}
  Y.~Hikida and S.J.~Rey,
  ``Can branes travel beyond CTC horizon in G\"odel universe?,''
  Nucl.\ Phys.\ {\bf B669} (2003) 57
  [hep-th/0306148].

\bibitem{Drukker:2003mg}
  N.~Drukker, B.~Fiol and J.~Simon,
  ``G\"odel type universes and the Landau problem,''
  JCAP {\bf 0410} (2004) 012
  [hep-th/0309199].
 
\bibitem{Israel:2003cx}
  D.~Israel,
  ``Quantization of heterotic strings in a G\"odel/anti-de Sitter spacetime and chronology protection,''
  JHEP {\bf 0401} (2004) 042
  [hep-th/0310158].
 
\bibitem{Israel:2004vv}
  D.~Israel, C.~Kounnas, D.~Orlando and P.M.~Petropoulos,
  ``Electric/magnetic deformations of $S^3$ and AdS$_3$, and geometric cosets,''
  Fortsch.\ Phys.\  {\bf 53} (2005) 73
  [hep-th/0405213].

\bibitem{Israel:2004cd}
  D.~Israel, C.~Kounnas, D.~Orlando and P.M.~Petropoulos,
  ``Heterotic strings on homogeneous spaces,''
  Fortsch.\ Phys.\  {\bf 53} (2005) 1030
  [hep-th/0412220].

\bibitem{Mazur:1986gb}
  P.O.~Mazur,
  ``Spinning cosmic strings and quantization of energy,''
  Phys.\ Rev.\ Lett.\  {\bf 57} (1986) 929.

\bibitem{Landau}
E.M. Lifshitz and L. P. Pitaevski, \textsl{Physical kinetics}, Pergamon Press, Oxford, 1981, p.254.

\bibitem{Sonner}
  J.~Sonner,
  ``A rotating holographic superconductor,''
  Phys.\ Rev.\ {\bf D80} (2009) 084031
  [arXiv:0903.0627 [hep-th]].

\bibitem{Kovtun2}
  S.A.~Hartnoll and P.~Kovtun,
  ``Hall conductivity from dyonic black holes,''
  Phys.\ Rev.\ {\bf D76} (2007) 066001
  [arXiv:0704.1160 [hep-th]].

\bibitem{Avron:1995fg} 
  J.E.~Avron, R.~Seiler and P.G.~Zograf,
  ``Viscosity of quantum Hall fluids,''
  Phys.\ Rev.\ Lett.\  {\bf 75} 697 (1995).

\bibitem{Saremi:2011ab} 
  O.~Saremi and D.T.~Son,
  JHEP {\bf 1204}, 091 (2012)
  [arXiv:1103.4851 [hep-th]].

\bibitem{Nigel}

N.R. Cooper, N.K. Wilkin, and J.M.F. Gunn, ``Quantum phases of vortices in rotating Bose--Einstein condensates,'' Phys. Rev. Lett. {\bf 87} 120405 (2001).


\bibitem{LPP_new}
R. G. Leigh, A. C. Petkou and P. M. Petropoulos, to appear.

\bibitem{CLPPPS}
M. Caldarelli, R.G. Leigh, A.C. Petkou, P.M. Petropoulos, V. Pozzoli and K. Siampos, to appear.

\end{thebibliography}
\end{document}